\pdfoutput=1
\documentclass[11pt,twoside,a4paper,cmspaper,final,collab]{cms-tdr}
\def\svnVersion{4a81c8b}\def\svnDate{2025/11/16}\def\cmsCernNoTag{CERN-EP-2025-217}\def\cmsCernDate{\today}\def\cmsMessage{Submitted to Science Bulletin}
\begin{document}\cmsNoteHeader{B2G-24-007}

\newcommand{\cmsTable}[1]{\resizebox{\textwidth}{!}{#1}}
\newlength\cmsTabSkip\setlength{\cmsTabSkip}{1ex}
\newcommand{\mj}{\ensuremath{m_{\text{j}}}\xspace}
\newcommand{\mjg}{\ensuremath{m_{\text{j}\PGg}}\xspace}
\newcommand{\Zjets}{\ensuremath{\PZ\text{+jets}}\xspace}
\DeclareRobustCommand{\PS}{{\HepParticle{S}{}{}}\xspace}
\renewcommand{\PX}{\ensuremath{\HepParticle{X}{}{}}\xspace}
\newcommand{\mZp}{\ensuremath{m_{\PZpr}\xspace}}
\newcommand{\mS}{\ensuremath{m_{\rm\PS}\xspace}}
\cmsNoteHeader{B2G-24-007}

\title{Search for heavy \texorpdfstring{$\PH\PGg$}{H gamma} and \texorpdfstring{$\PZ\PGg$}{Z gamma} resonances with a bottom quark-antiquark pair in the final state in proton-proton collisions at \texorpdfstring{$\sqrt{s}=13\TeV$}{sqrt(s) = 13 TeV} }

\date{\today}
\abstract{A search for heavy resonances decaying into a Higgs boson (\PH) or a \PZ boson and a photon (\PGg), with the \PH or \PZ bosons decaying to a bottom quark-antiquark pair (\bbbar) is presented. The analysis is performed using proton-proton collision data at $\sqrt{s} = 13\TeV$ collected by the CMS experiment at the CERN Large Hadron Collider, corresponding to an integrated luminosity of 138\fbinv. The analyzed events contain a photon and a massive large-radius jet originating from a Lorentz-boosted \bbbar system. An advanced transformer-based algorithm classifies jets according to their substructure and quark flavors, forming a tagger that identifies jets as candidates from $\PH/\PZ\to\bbbar$ decays. A set of parametric functions is used to fit the photon-jet invariant mass spectrum and to extract potential signals. No significant excess is observed above the standard model expectations. The results set upper limits at 95\% confidence level on the product of the cross section and the branching fraction for spin-1 $\PH\PGg$ resonances and spin-0 $\PZ\PGg$ resonances, below 0.1 and 0.3\unit{fb}, respectively, representing the most stringent limits to date.}

\hypersetup{
pdfauthor={CMS Collaboration},
pdftitle={Search for heavy H gamma and Z gamma resonances with a b quark pair final state in proton-proton collisions at sqrt(s) = 13 TeV},
pdfsubject={CMS},
pdfkeywords={CMS, Higgs, photon, b quarks pair, Particle Transformer, Global Particle Transformer, ParT, GloParT, Particle Net, PNet, regressed mass, Tagger, resonance}
}

\maketitle 

\section{Introduction} \label{Sec:Intro}

Exploring extensions of the standard model (SM) is one of the main objectives of the CERN Large Hadron Collider (LHC) research program.
Physics beyond the SM (BSM) may reveal itself through the production of new heavy particles that decay into SM particles.
Numerous searches for massive resonances decaying into pairs of SM particles, such as dijet or diboson analyses~\cite{CMS:2021itu,CMS:2022pjv,CMS:2021klu,CMS:2019gwf,CMS:2024hbn,CMS:2021xor,CMS:2021fyk,CMS:2024phk,CMS:2017dyb,CMS:2018cno,CMS:2021zxu,CMS:2024nht,
ATLAS:2020qiz,ATLAS:2020fry,ATLAS:2019nat,ATLAS:2024qvg,ATLAS:2019nat,ATLAS:2020fry,ATLAS:2022zuc,ATLAS:2020qiz, ATLAS:2022enb,ATLAS:2019fgd}, have been carried out with the goal of uncovering any possible hints of BSM physics.

This paper presents a search for resonances that decay into a photon (\PGg) and either a Higgs boson (\PH) or a \PZ boson, with the \PH or \PZ boson subsequently decaying into a bottom quark-antiquark pair (\bbbar). 
In particular, this analysis considers a narrow-width spin-1 resonance, denoted as $\PZpr$, decaying as $\PZpr\to\PH\PGg\to\PQb\PAQb\PGg$, as well as a narrow- or broad-width spin-0 resonance, $\PS$, decaying as $\PS\to\PZ\PGg\to\PQb\PAQb\PGg$. 
The analysis focuses on heavy $\PZpr$ and $\PS$ resonances that decay into a \PGg and a Lorentz-boosted $\PQb\PAQb$ quark system. 
The hadronization products of the $\PQb\PAQb$ system form a single, massive, large-radius jet (j) with two subjets, typically featuring displaced secondary vertices, which are characteristic of the hadronization of \PQb quarks.

The \PZpr resonant signal is predicted by several BSM scenarios, such as the examples studied in Ref.~\cite{Dobrescu:2017sue}. 
The one-loop $\PZpr\to\PH\PGg$ process discussed in that paper is used as the benchmark signal model in this analysis, as in earlier studies~\cite{CMS:2018cno,ATLAS:2020jeb}.
The \PZpr resonance is a neutral spin-1 gauge boson associated with an additional U(1)' symmetry that extends the SM electroweak sector, and because its decay to H$\gamma$ is assumed to proceed through a heavy-quark loop, the corresponding branching fraction is typically $10^{-5}$ or smaller. 
Nonetheless, these models predict scenarios in which a significant portion of the parameter space leaves open the possibility that the $\PZpr\to\PH\PGg$ mode could be observed, owing to the small SM background.
In this analysis, the fractional width of the \PZpr boson, $\Gamma_{\PZpr}/\mZp$, where $\mZp$ ($\Gamma_{\mZp}$) is the \PZpr boson mass (intrinsic width), is assumed to be narrow and the search is conducted for signal masses in the range of 0.7--3.5\TeV.

Neutral massive bosons with spin 0 or 2 can also decay via the $\PZ\PGg$ channel.
Such decays are predicted in various BSM extensions, including technicolor~\cite{EL} and little Higgs~\cite{FS} models, as well as models with extended Higgs sectors~\cite{BT,LLS} and extra spatial dimensions~\cite{DHR,ASS}.
In this analysis, the $\PZ\PGg$ signal considered is a heavy scalar boson, denoted in this paper as \PS, similar to the SM Higgs boson. This model has been used as a benchmark in similar analyses, including Refs.~\cite{CMS:2017dyb,PhysRevD.98.032015}.
Although the $\PZ\PGg$ channel results are interpreted for scalar resonances, they are approximately applicable to spin-2 states, as the signal acceptance is only weakly dependent on the resonance's spin~\cite{Aaboud:2017uhw}.
For the $\PS\to\PZ\PGg$ process, three width scenarios of $\Gamma_{\PS}/\mS$ are considered, with signal hypotheses in the \mS~range 0.7--3.5 \TeV.

\begin{figure}[!htbp] \centering
\includegraphics[width=0.8\linewidth]{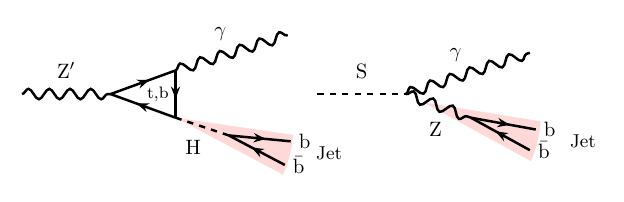}
\caption{Schematic illustration indicating the topology of the signal processes under consideration in this search. 
The spin-1 $\PZpr\to\PH\PGg\to\bbbar\PGg$ signal scenario is shown on the left, and the spin-0 $\PS\to\PZ\PGg\to\bbbar\PGg$ signal scenario is shown on the right.}
\label{fig:diagram}
\end{figure}

In both the $\PZpr\to\PH\PGg$ and $\PS\to\PZ\PGg$ signal scenarios, the $\PH$ or $\PZ$ boson can decay hadronically into a \bbbar system. 
The branching fraction $\mathcal{B}(\PH \to \bbbar)\approx58\%$ is the largest among all Higgs boson decay modes, while the \PZ boson branching fraction $\mathcal{B}(\PZ\to\bbbar)\approx15\%$ is also a sizable \PZ boson decay mode. 
Since $\mZp,\mS>0.7\TeV$, the \bbbar system typically receives a significant Lorentz boost, and its hadronization products can be reconstructed as a single, large-radius jet. Figure \ref{fig:diagram} displays schematic diagrams of the signal processes probed.

The analysis employs advanced machine-learning (ML) techniques to select $\PH\to\bbbar$ and $\PZ\to\bbbar$ candidate jets. 
For the identification of candidate jets based on their reconstructed mass, an ML-based jet mass variable is constructed using the \textsc{ParticleNet} algorithm~\cite{Qu:2019gqs}, which has been shown to improve the jet mass resolution by approximately 20\%~\cite{cms2021mass} compared to the previously used soft-drop mass technique~\cite{Larkoski:2014wba} for jets formed by the $\PH\to\bbbar$ decays.
The substructure of candidate jets is analyzed using the state-of-the-art global particle transformer (\textsc{GloParT}) algorithm~\cite{Qu:2022mxj}, which provides discriminant variables that efficiently suppress background processes, leading to a significantly improved identification of jets with the \bbbar structure.

This analysis uses proton-proton ($\Pp\Pp$) collision data collected by the CMS experiment at $\sqrt{s} = 13\TeV$ in 2016--2018, corresponding to an integrated luminosity of 138\fbinv~\cite{CMS-LUM-17-003,CMS-PAS-LUM-17-004,CMS-PAS-LUM-18-002}.
The data are divided into four signal regions (SRs) based on the jet mass and the \bbbar tagger score. 
The invariant mass of the $\PGg$+jet system (\mjg) is studied in each SR for the presence of signal, which manifests as a peak in the \mjg spectrum on top of a smoothly falling background.
While this search is optimized for $\PH\PGg$ and $\PZ\PGg$ resonant signals, it is also moderately sensitive to other resonances ($\text{X}$) not specifically targeted by the analysis, including processes involving photon pairs, such as $\text{X}\to\PH\text{A}\to\bbbar\PGg\PGg$ and $\text{X}\to\PZ\text{A}\to\bbbar\PGg\PGg$ (with $\text{A}$ representing a light hypothetical (pseudo)scalar particle).
Indeed, for $\text{A}$ masses at the \GeV scale or below, the two photons would be collimated and reconstructed as a single photon, yielding the same signature as in this analysis.

This paper is organized as follows.
Section~\ref{sec:cms} describes the CMS detector.
The background Monte Carlo (MC) simulated samples and the signal modeling are discussed in Section~\ref{sec:bkg_simulation}.
The triggers employed for data collection and the event reconstruction are described in Section~\ref{sec:Objects}.
Event selection and categorization are presented in Section~\ref{sec:event_selection}.
Section~\ref{sec:bkg_estimation} describes the estimation of the SM background.
Systematic uncertainties are presented in Section~\ref{sec:uncertainties}.
The results and their interpretation are given in Section~\ref{sec:Results}.
The paper is summarized in Section~\ref{sec:sum}.
Tabulated results for this analysis are provided in the HEPData record~\cite{hepdata}.

\section{The CMS detector} \label{sec:cms}

The CMS apparatus~\cite{CMS:2008xjf,CMS:2023gfb} is a multipurpose, nearly hermetic detector, designed to trigger on~\cite{CMS:2020cmk,CMS:2016ngn,CMS:2024aqx} and identify electrons, muons, photons, and (charged and neutral) hadrons~\cite{CMS:2020uim,CMS:2018rym,CMS:2014pgm}. Its central feature is a superconducting solenoid of 6\unit{m} internal diameter, providing a magnetic field of 3.8\unit{T}. Within the solenoid volume are a silicon pixel and a strip tracker, a lead tungstate crystal electromagnetic calorimeter (ECAL), and a brass and scintillator hadron calorimeter (HCAL), each composed of a barrel and two endcap sections. Forward calorimeters extend the pseudorapidity ($\eta$) coverage provided by the barrel and endcap detectors. Muons are reconstructed using gas-ionization detectors embedded in the steel flux-return yoke outside the solenoid. More detailed descriptions of the CMS detector, together with a definition of the coordinate system used and the relevant kinematic variables, can be found in Refs.~\cite{CMS:2008xjf,CMS:2023gfb}. 

The ECAL consists of 75\,848 PbWO$_4$ crystals, each with a transverse dimension approximately matching the Moli\`ere radius of the material and 25 radiation lengths in depth.
The ECAL barrel section covers the region $\abs{\eta}<1.48$, while the two endcap sections extend over $1.48<\abs{\eta}<3.00$. In the barrel section, an energy resolution of about 1\% is achieved for unconverted or late-converting photons with energies of a few tens of \GeV.
The energy resolution of the remaining, converted barrel photons is about 1.3\% up to $\abs{\eta} = 1$, changing to about 2.5\% at $\abs{\eta} = 1.4$.
In the endcaps, the energy resolution is about 2.5\% for unconverted or late-converting photons, and between 3 and 4\% for converted photons~\cite{CMS:2020uim,CMS:2015myp}.
The energy deposits in ECAL and HCAL cells are summed to define the calorimeter tower energies, which are subsequently used to provide the energies and directions of hadronic jets.
The silicon tracker measures charged particles within the pseudorapidity range $\abs{\eta} < 3.0$. During the LHC running period when the data used in this Letter were recorded, the silicon tracker consisted of 1856 silicon pixel and 15\,148 silicon strip detector modules. Details on the pixel detector can be found in Ref.~\cite{Phase1Pixel}. For nonisolated particles of transverse momentum (\pt) in 1--10\GeV range and $\abs{\eta} < 3.0$, the track resolutions are typically 1.5\% in \pt and 20--75\mum in the transverse impact parameter~\cite{DP-2020-049}

Events of interest are selected using a two-tiered trigger system.
The first level, composed of custom hardware processors, uses information from the calorimeters and muon detectors to select events at a rate of around 100\unit{kHz} within a fixed latency of 4\mus~\cite{CMS:2020cmk}.
The second level, known as the high-level trigger, consists of a farm of processors running a version of the full event reconstruction software optimized for fast processing, and reduces the event rate to a few \unit{kHz} before data storage~\cite{CMS:2016ngn,CMS:2024aqx}.

\section{Simulated samples and signal modeling} \label{sec:bkg_simulation}

The selection optimization is performed using simulated signal and SM background samples.
These include events from quantum chromodynamics (QCD) multijet production, \PW and \PZ boson production, and \ttbar production, each accompanied by an additional photon.
Additional samples of the same processes without a photon are also considered, accounting for cases where hadrons or electrons may be misidentified as photons. 
The overlap between processes with and without photons is resolved by applying a generator-level disambiguation procedure. For example, in the QCD process, only events without prompt photons---collectively referred to as nonprompt---are retained, while for the QCD+$\PGg$ process, only events with a prompt photon are kept.
Events for all processes are generated at leading order (LO) or next-to-LO (NLO) precision using the \MGvATNLO~2.4.2 generator~\cite{aMCNLO}.

For both signal and background samples, fragmentation and hadronization are simulated with \PYTHIA 8.230~\cite{Sjostrand:2014zea}.
The underlying event is modeled using the CP5 tune~\cite{Sirunyan:2019dfx}, and \GEANTfour~\cite{GEANT4:2002zbu} is used to simulate the detector response.
Additional $\Pp\Pp$ interactions occurring in the same or nearby bunch crossings, known as pileup (PU), are overlaid on the main collision during simulation.
The NNPDF 3.1 next-to-NLO parton distribution function (PDF) set is used for simulating the hard processes.

To account for residual differences between the distributions of the multiplicity of PU vertices simulated in MC samples and those observed in data, a reweighting of MC events is performed.
Photon identification efficiency and energy resolution are corrected for differences between data and simulation, as measured in $\PZ\to\Pep\Pem$ events~\cite{CMS:2015myp}, since photons and electrons are subject to similar reconstruction algorithms.

Simulated spin-1 \PZpr resonance signal samples with the \PZpr boson decaying to $\PH\PGg$ are generated at LO based on the benchmark model proposed in Ref.~\cite{Dobrescu:2017sue}. 
Following the discussion in Ref.~\cite{Dobrescu:2017sue}, the intrinsic fractional width of the \PZpr signal, $\Gamma_{\PZpr}/\mZp$, is set to 0.01\%, which is negligible compared to the detector resolution effects.

The model used to simulate the $\PZ\PGg$ resonance signal is a spin-0 heavy Higgs boson ($\PS$) generated using \PYTHIA at LO.
Three resonance width hypotheses are considered, with fractional widths $\Gamma_{\PS}/\mS$ of 0.014, 5.6, and 10\%, as has been done in previous analyses, allowing for a direct comparison.
The simulation assumes no interference between the signal and the SM nonresonant $\PZ\PGg$ production.

For both \PZpr and $\PS$ boson signal scenarios and all width hypotheses, fully simulated signal samples are generated for 14 resonance masses.  
These are produced across the mass range 700--3500\GeV, with step sizes of 100\GeV for 700--1000\GeV, 200\GeV for 1000--2600\GeV, and the final point at 3500\GeV. 
The $\mjg$ spectra of the signal samples are modeled using a bifurcated double-sided Crystal Ball (DCB) function~\cite{CrystalBall,Gaiser:1982yw}, which provides a good description of the simulated distributions for all signal points.
For the masses between the 14 simulated \mZp and \mS points, the signal \mjg shape is obtained by smoothly interpolating the DCB function parameters using a cubic spline fit~\cite{Read:1999kh}.
The signal modeling obtained from interpolation is validated through comparisons across several simulated signal samples.
For the broad-width signal scenarios ($\Gamma_{\PS}/\mS = 5.6,~10\%$), a significant fraction of the signal events are produced off-shell, spreading across the \mjg spectrum at values lower than the nominal \mS, as illustrated in the right plot of Fig.~\ref{fig:PreselectionPlots}. To ensure that the DCB model accurately describes the signal for broad widths, only the on-shell part of the \mjg distribution is considered, \ie, the fraction of signal events within the $\mS\pm0.33\mS$ range. 
This on-shell portion of the signal, which is consistently well modeled by the DCB function across all scenarios, is employed in the fit and in the interpretation of the results.
Modeling the full off-shell tail would require a more complex function and offers little sensitivity gain, as its shape resembles the background and is absorbed by the fit.
The systematic uncertainties are also interpolated with a cubic spline to ensure a smooth evolution across all tested signal scenarios.

\section{Event reconstruction} \label{sec:Objects}

The primary interaction vertex (PV) is taken to be the vertex corresponding to the hardest scattering in the event, evaluated using tracking information alone, as described in Section 9.4.1 of Ref.~\cite{CMS-TDR-15-02} and Ref.~\cite{CMS:2020ebo}.

A particle-flow (PF) algorithm~\cite{CMS:2017yfk} aims to reconstruct and identify each interacting particle in an event, with an optimized combination of information from the various elements of the CMS detector.
The energy of photons is obtained from the ECAL measurement. 
The energy of electrons is determined from a combination of the track momentum at the PV as determined by the tracker, the corresponding ECAL cluster energy, and the energy sum of all bremsstrahlung photons spatially compatible with originating from the electron track.
The energy of muons is obtained from the corresponding track momentum. 
The energy of charged hadrons is determined from a combination of their momentum measured in the tracker and the matching ECAL and HCAL energy deposits, corrected for the response function of the calorimeters to hadronic showers.
Finally, the energy of neutral hadrons is obtained from the corresponding corrected ECAL and HCAL energies.

For each event, hadronic jets are clustered from these reconstructed particles (``PF candidates'') using the infrared- and collinear-safe anti-\kt jet clustering algorithm~\cite{Cacciari:2008gp,Cacciari:2011ma}, as implemented in the \textsc{FastJet} package~\cite{Cacciari:2011ma}.
The clustering algorithm is run twice over the same inputs with the distance parameters 0.4 and 0.8, resulting in AK4 and AK8 jet collections, respectively.
Jet momentum is determined as the vectorial sum of all particle momenta in the jet, and is found from simulation to be, on average, within 5 to 10\% of the true momentum over the entire \pt spectrum and detector acceptance~\cite{CMS:2016lmd}.

Particles from PU interactions can contribute additional tracks and calorimetric energy depositions within the same or nearby bunch crossings, increasing the apparent jet momentum.
The PU-per-particle identification algorithm~\cite{CMS:2020ebo,Bertolini:2014bba} is used to mitigate the effect of PU in AK8 jets at the reconstructed particle level, making use of local shape information, event PU properties, and tracking information.
A local shape variable is defined, which distinguishes between collinear and soft diffuse distributions of other particles surrounding the particle under consideration. 
The former is attributed to particles originating from the hard scatter, and the latter to particles originating from PU interactions. 
Charged particles identified to be originating from PU vertices are discarded. 
For each neutral particle, the local shape variable is computed using the surrounding charged particles compatible with the PV within the tracker acceptance ($\abs{\eta}<2.5$), and using both charged and neutral particles in the region outside of the tracker coverage. 
The momenta of the neutral particles are then rescaled according to their probability to originate from the PV deduced from the local shape variable, thus superseding the need for jet-based PU corrections~\cite{CMS:2020ebo}.

Jet energy corrections are derived from simulation to bring the measured response of jets to that of particle level jets on average.
Measurements in data of the momentum balance in dijet, \PGg{}+$\text{jet}$, \Zjets, and multijet events are used to account for any residual differences in the jet energy scale (JES) between data and simulation~\cite{CMS:2016lmd}.
When combining information from the entire detector, the jet energy resolution (JER) typically amounts to 15--20\% at 30\GeV, 10\% at 100\GeV, and 5\% at 1\TeV~\cite{CMS:2016lmd}.
Additional selection criteria are applied to each jet to remove jets potentially dominated by anomalous contributions from various subdetector components or reconstruction failures~\cite{CMS-PAS-JME-16-003}.

The AK4 jets originating from the hadronization of \PQb-quarks are identified as ``\PQb-tagged'' using the ML-based \textsc{DeepJet}~\cite{Bols:2020bkb,CMS-DP-2018-058} algorithm that uses tracks within the jet that are displaced from the PV, identified secondary vertices, jet kinematic variables, neutral PF candidates, and information related to the presence of soft leptons in the jet as inputs.
A threshold on the \textsc{DeepJet} score is used, corresponding to a 1\% probability of misidentifying a light-flavor quark or a gluon ($\PQu\PQd\PQs\Pg$) AK4 jet with $\pt > 30\GeV$ and $\abs{\eta}<2.5$ as a \PQb quark jet, while maintaining approximately 83\% signal efficiency.

The missing transverse momentum vector, \ptvecmiss, is computed as the negative vector sum of the transverse momenta ($\Vec{\pt}$) of all the PF candidates in an event, and its magnitude is denoted as \ptmiss~\cite{CMS:2019ctu}.
The \ptvecmiss is modified to account for corrections to the energy scale of the reconstructed jets in the event.
Anomalous high-\ptmiss events can occur as a result of a variety of reconstruction failures, detector malfunctions, or noncollision backgrounds.
Such events are rejected by criteria designed to identify more than 85--90\% of spurious high-\ptmiss events with a signal efficiency loss of less than 0.1\%~\cite{CMS:2019ctu}.

The AK8 jets arising from hadronic decays of \PH and \PZ bosons are identified using the \textsc{GloParT} ML algorithm~\cite{Qu:2022mxj}, which uses PF candidates and secondary vertices as inputs. 
The algorithm is trained to classify jets into a total of 314 distinct classes, and its output consists of a score for each class corresponding to the probability that the jet belongs to that class. 
This large jet classification model extends beyond the needs of this analysis, offering a comprehensive set of jet classes designed to support the broad and diverse CMS data analysis program.
It leverages larger training data sets and more complex classification tasks to enhance the overall tagging performance, as demonstrated in Ref.~\cite{Li:2024htp}.
The present analysis employs the mass-decorrelated implementation~\cite{CMS-DP-2020-002} of \textsc{GloParT}.
The identification of \PH and \PZ bosons uses the $\text{Xbb}$ score, defined as the ratio of the $\PX\to\bbbar$ class score to the sum of five scores of categories corresponding to jets produced purely via QCD multijet processes and $\PX\to\bbbar$ jets.
The \textsc{ParticleNet} algorithm also outputs an estimate of the jet's mass, referred to as the regressed jet mass, $m_\text{j}$, which is used in this analysis for further discrimination between \bbbar-candidate jets and background jets.

Photon candidates are reconstructed from energy deposits in the ECAL, referred to as superclusters, using a clustering algorithm that groups deposits consistent with the expected electromagnetic shower shape and accounts for energy recovery from photon conversions and electron bremsstrahlung. Since the dominant background arises from prompt-photon production rather than misidentified photons, a ``loose'' identification working point~\cite{CMS:2020uim} is used to maximize photon efficiency, albeit with a slightly higher misidentification rate. 
 The identification relies on photon shower shape and isolation variables, computed from PF candidates within a cone of radius $\Delta R = \sqrt{\smash[b]{(\Delta\eta)^2+(\Delta\phi)^2}} = 0.3$ around the photon, where $\phi$ is the azimuthal angle, with corrections applied for PU effects. 
Additionally, a conversion-safe electron veto~\cite{CMS:2020uim} is applied. 
Further details on the reconstruction and identification can be found in Ref.~\cite{CMS:2020uim}. 
The loose working point achieves approximately 90\% efficiency, with weak \pt dependence, while reducing misidentified photon backgrounds by a factor of 7 and limiting their contribution to below 7\% of the total background.
Photon candidates are further required to satisfy additional criteria to suppress backgrounds from jets and electrons, using electromagnetic shower shape and isolation variables, hadronic activity, and the absence of a matching track, as detailed in Ref.~\cite{CMS:2024nht}.

In order to reject SM background events with prompt electrons and muons (collectively referred to as leptons), selection criteria are applied in order to distinguish isolated leptons from leptons arising from heavy-flavor decays. Isolated electron (muon) candidates are required to be within the geometrical acceptance, which is $\abs{\eta}<2.5~(2.4)$, and have $\pt>35~(20)~\GeV$. Identified isolated electrons and muons are then selected based on the nearby hadronic activity in the event using an isolation variable defined by summing the \pt of all PF candidates in a cone of radius ${\Delta R = 0.3\,(0.4)}$ around the electron (muon) track, employing a correction for the contribution of neutral particles from PU interactions. More details on lepton isolation can be found in Refs.~\cite{CMS:2020uim,CMS:2018rym}.

\section{Event selection}\label{sec:event_selection}

\subsection{Trigger}\label{subsec:Trigger}

The data sample used for this analysis is collected using a set of single-photon triggers that select events with at least one \Pg candidate within $\abs{\eta}<2.5$ and passing photon \pt thresholds of 165 or 175 \GeV for data collected in 2016 and 200 \GeV for 2017–2018.
The efficiencies of these triggers are measured using unbiased data samples collected using triggers that select events with single muons. 
These single-photon triggers have an efficiency above 98\% relative to the offline selection criteria described below, and are consistent between data and simulation across the full mass range and data set.
Residual inefficiencies are taken into account as a systematic uncertainty, as discussed in Section~\ref{sec:uncertainties}.
This choice of triggers is identical to that used in Refs.~\cite{CMS:2018cno,CMS:2021zxu}, where further details can be found.

\subsection{Event preselection} \label{subsec:Preselection_and_SR}

Events considered in this analysis are selected in two steps. 
The first step, referred to as preselection, applies a loose set of requirements for events with roughly signal-like kinematic features, and the resulting sample is used in optimization studies.
A more stringent set of requirements, optimized for high sensitivity in the SRs, defines the sample used in the statistical analysis.
The analysis focuses on AK8 jets from \PH or \PZ boson decays, hereafter referred to simply as ``jets'', while their regressed mass estimator, \mj, is likewise referred to as the jet mass. 

The preselection requires exactly one photon in the region $\abs{\eta_\PGg}<1.44$ or $1.57<\abs{\eta_\PGg}<2.5$~\cite{CMS:2020uim} with $\smash{\pt^{\PGg}>225\GeV}$ that satisfies the identification conditions described in Section~\ref{sec:Objects}.
In addition, the preselection requires events to have at least one jet with $\pt^{\text{j}}>250\GeV$ and $\mj>30\GeV$ in the region $\abs{\eta_\text{j}}<2.4$.
For events with more than one selected jet (7\% of the preselection), the jet with the highest Xbb score is chosen as the \PH or \PZ candidate, resulting in a 5--10\% misidentification rate depending on the resonance mass. 
Events that contain at least one isolated lepton (as defined in Section~\ref{sec:Objects}) are rejected.
Photons in the event are required to be separated from the jet using the requirement $\Delta R(\text{j},\PGg) > 1.1$, which ensures minimal overlap between the photon and jet.

A subdominant background contribution arises from SM processes involving top quarks, single top, and \ttbar production produced in association with a photon (collectively referred to as ``Top+\PGg''). 
To suppress these backgrounds, events containing \PQb-tagged AK4 jets satisfying the requirement $\Delta R(\PQb,\text{j}_{\text{AK4}})>1.2$, indicating spatial separation from the candidate \bbbar jet, are rejected at the preselection level. 

\begin{figure}[htbp!]\centering
\includegraphics[width=0.495\linewidth]{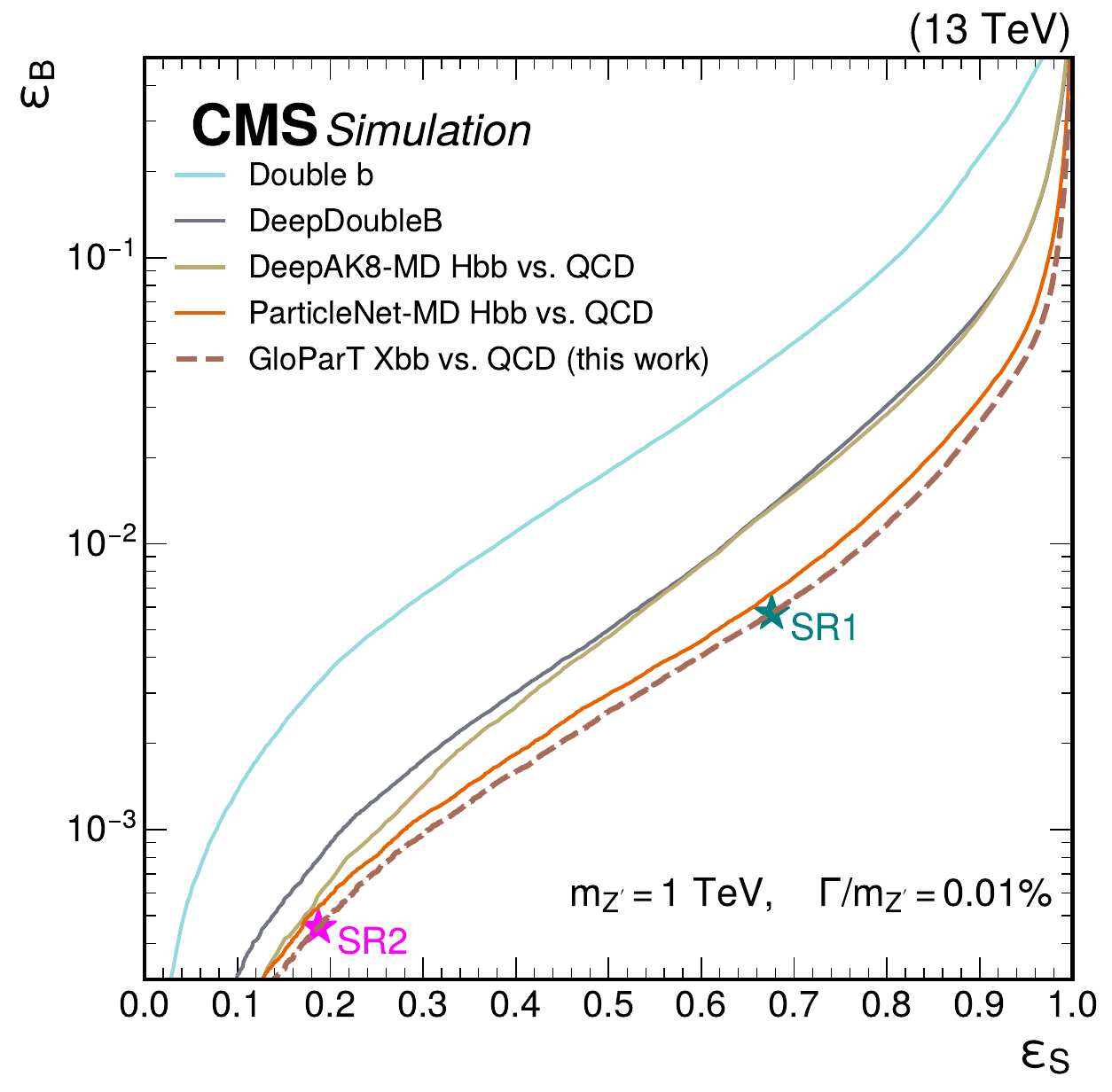}
\includegraphics[width=0.485\linewidth]{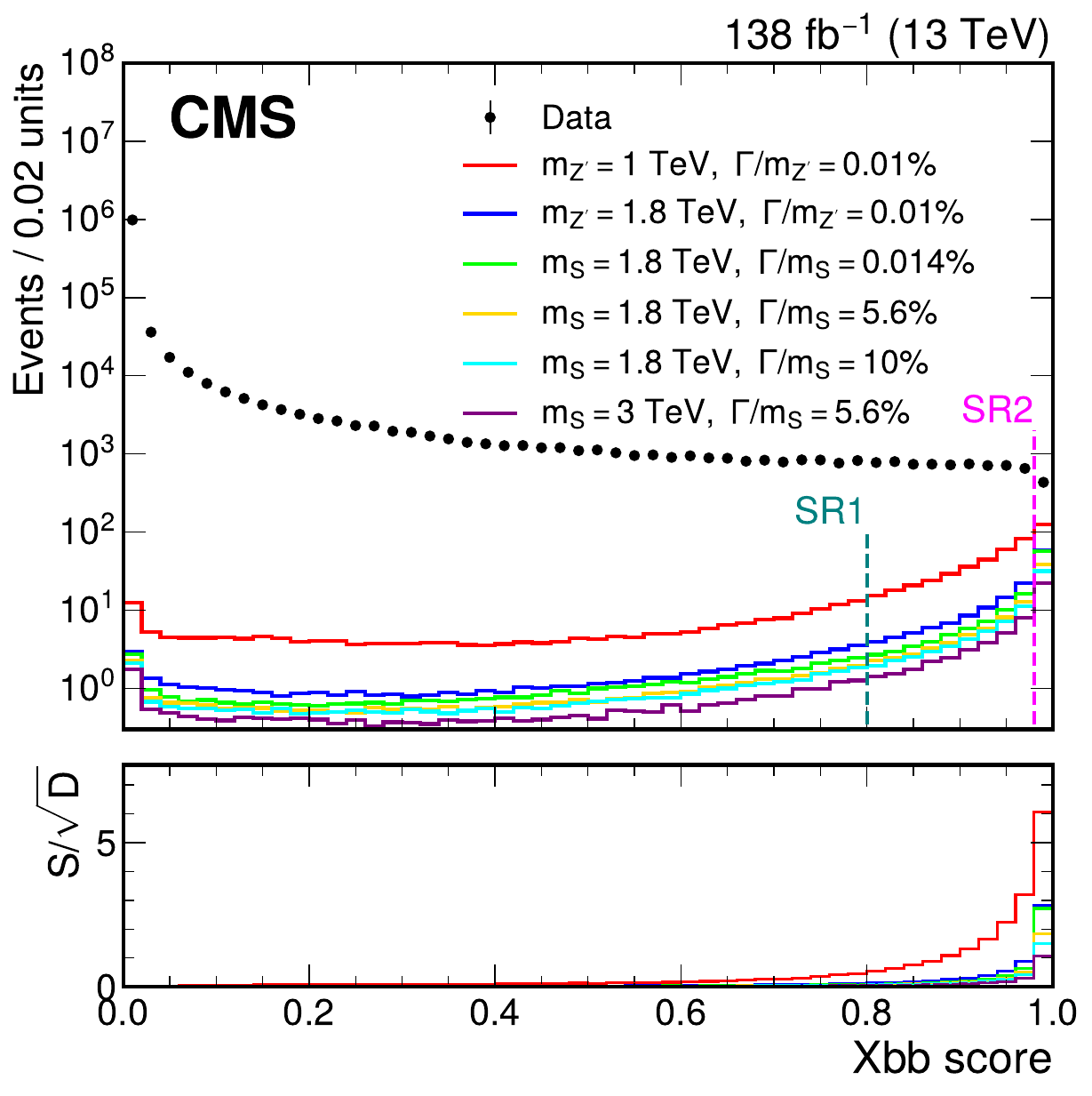}
\caption{The ROC curves of background efficiency ($\epsilon_\text{B}$) vs. signal efficiency ($\epsilon_\text{S}$) for the Xbb tagger and other taggers from previous analyses, evaluated using a $\mZp=1\TeV$ signal sample and SM background, are shown on the left. The taggers are listed in chronological order in the legend, based on their first use in similar searches (Refs.~\cite{CMS:2018cno,CMS:2020zge,CMS:2021roc,CMS:2022gjd}).
The right plot shows the Xbb score distributions for data and simulated signal samples, along with the significance metric, $S/\sqrt{D}$, as defined in the text.
Both plots use the preselection, with the \PZpr (\PS) signals normalized by setting $\sigma\mathcal{B}$ to the benchmark values (10~fb).}
\label{fig:TaggerPlots}
\end{figure}

The receiver operating characteristic (ROC) curves, shown on the left of Fig.~\ref{fig:TaggerPlots}, indicate the superior signal vs. background separation performance of the Xbb discriminator, as compared with other tagging algorithms used in previous CMS searches reported in Refs.~\cite{CMS:2018cno,CMS:2020zge,CMS:2021roc,CMS:2022gjd}. In the right of Fig.~\ref{fig:TaggerPlots}, the Xbb discriminant scores in simulated signal samples and the data are shown, along with the statistical significance metric $S/\sqrt{D}$ in the lower panel, where $S$ and $D$ are the signal and data event yields, respectively. 
Both plots in Fig.~\ref{fig:TaggerPlots} use the preselection criteria, with the \PZpr signals normalized to their benchmark cross sections times branching fraction ($\sigma\mathcal{B}$) values, while the \PS signals are  normalized to $\sigma\mathcal{B}=10$~fb.

\begin{figure}[htbp!]\centering
\includegraphics[width=0.495\linewidth]{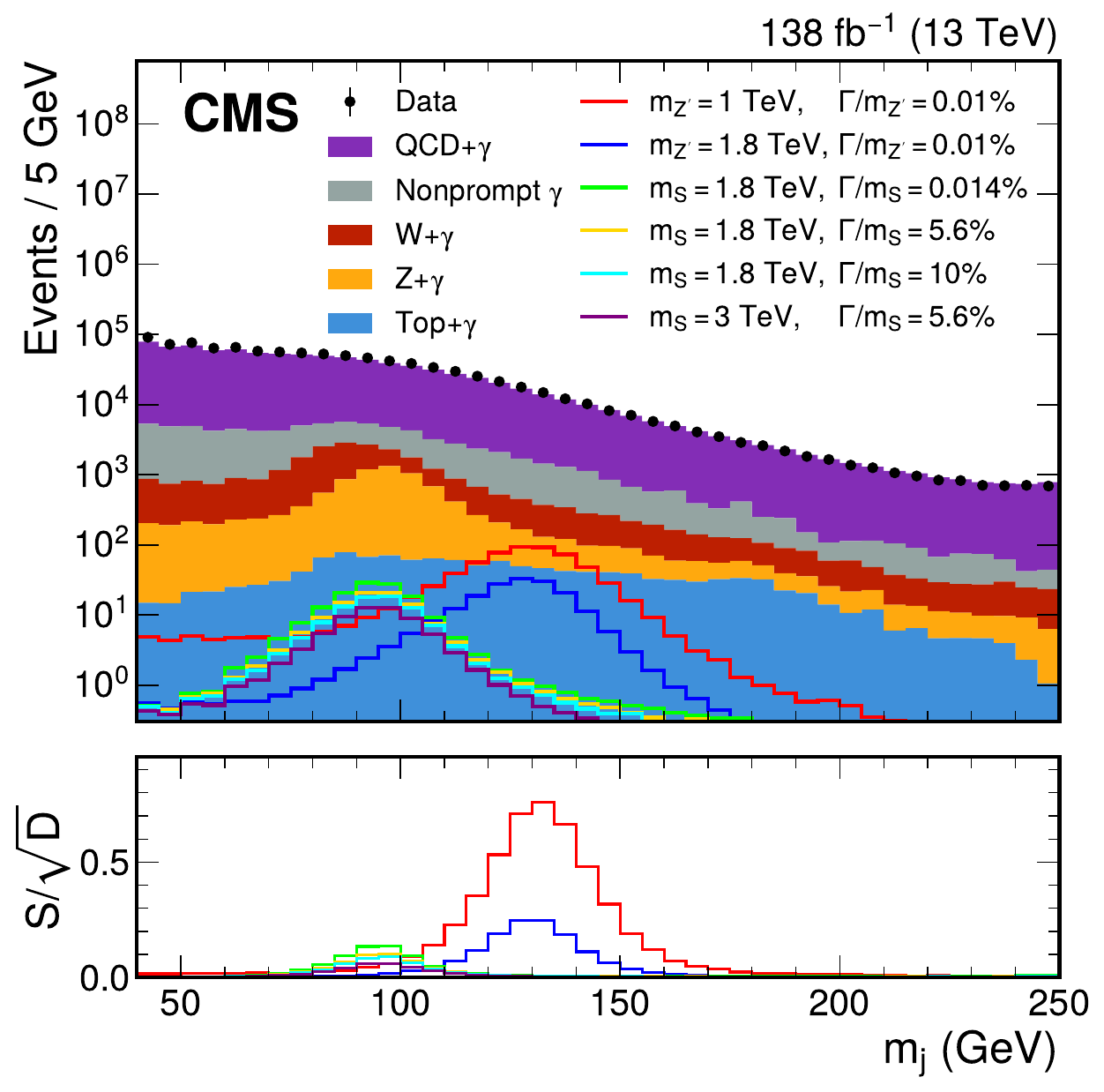}
\includegraphics[width=0.498\linewidth]{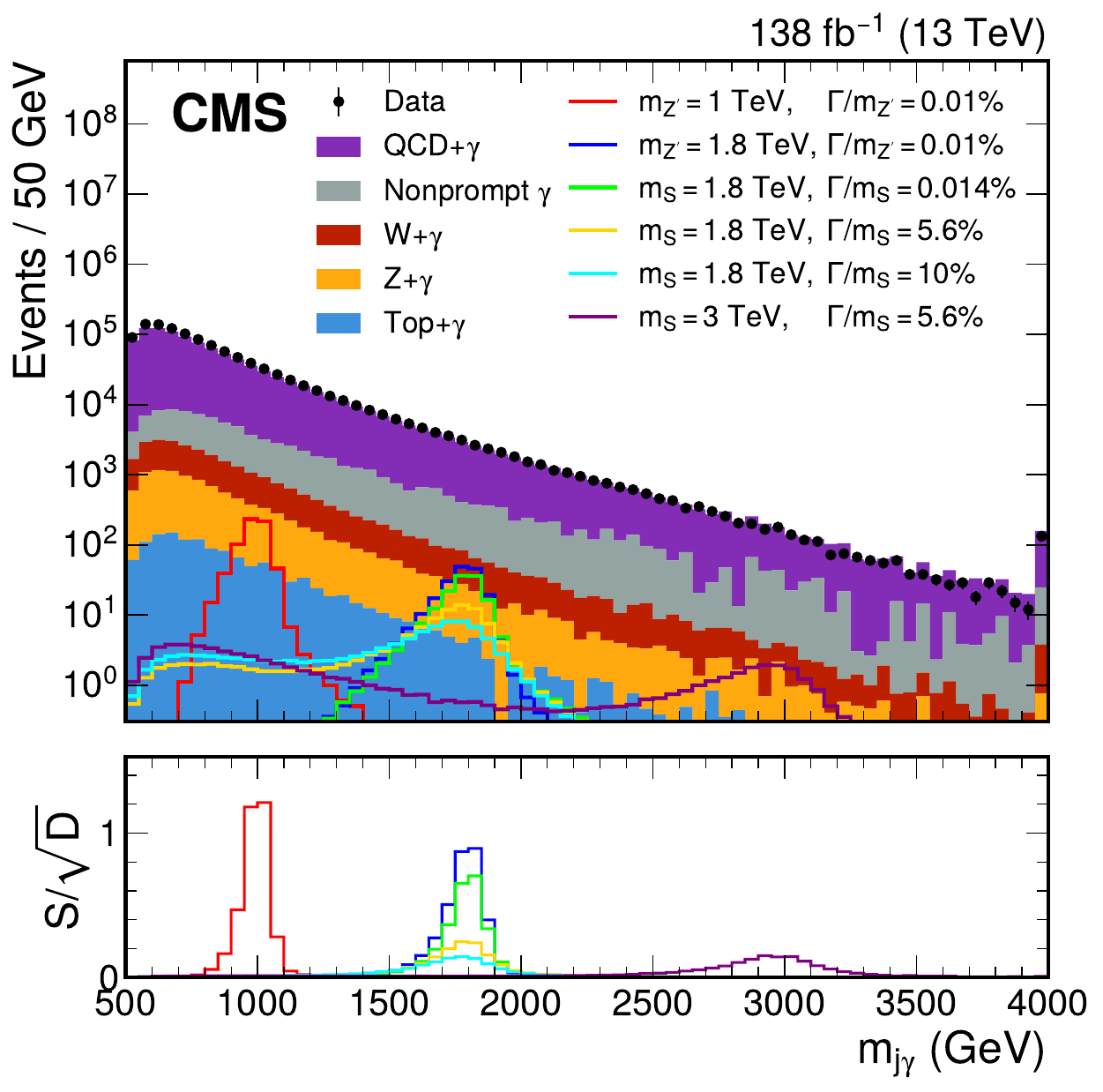}
\caption{The left and right plots present the distributions of \mj and \mjg, respectively, in data, MC background, and signal samples, with preselection criteria applied. 
Both plots use the preselection, with the \PZpr (\PS) signals normalized by setting $\sigma\mathcal{B}$ to the benchmark (10~fb) values.}
\label{fig:PreselectionPlots}
\end{figure}

The \mj and \mjg spectra at the preselection level are compared between data and simulated background samples in Fig.~\ref{fig:PreselectionPlots}. 
Six signal scenarios are overlaid, with four corresponding to the same mass, enabling a direct shape comparison among the simulated resonances. 
The lower panels show the $S/\sqrt{D}$ distributions.

\subsection{Signal region and event categorization} \label{subsec:selection_and_SR}

The full selection criteria, which define the SRs of this analysis, are a more stringent set of optimized event selection requirements that are applied in addition to the preselection criteria.

In order to further suppress the QCD multijet background, photons are required to lie within the ECAL barrel region, corresponding to $\abs{\eta_\PGg}<1.44$. 
This requirement takes advantage of the fact that signal events with masses above 700\GeV predominantly yield central photons, whereas photons produced in QCD background processes are more commonly observed in the forward regions of the detector.

\begin{figure}[htbp]\centering
\includegraphics[width=0.48\linewidth]{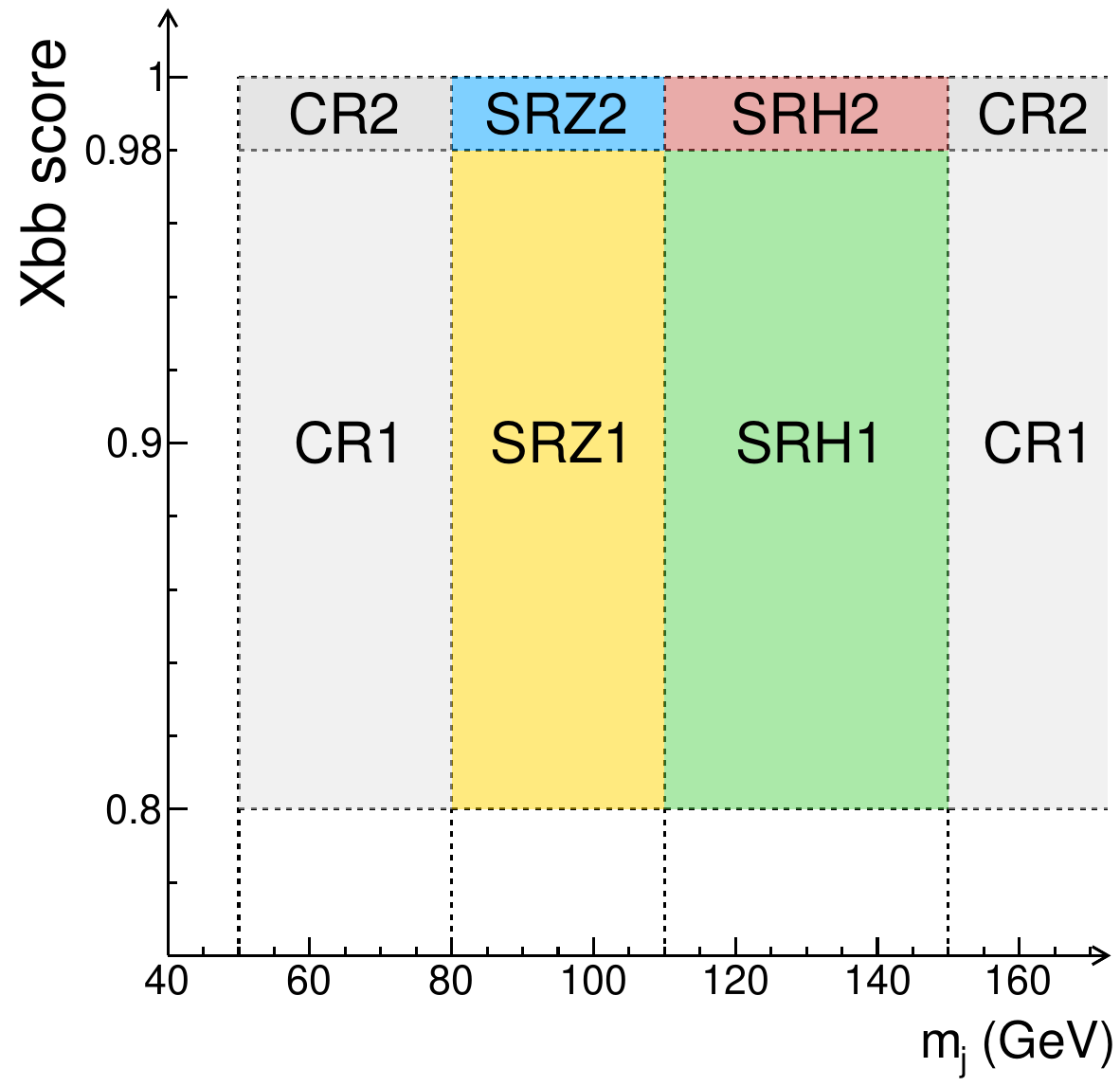}
\includegraphics[width=0.49\linewidth]{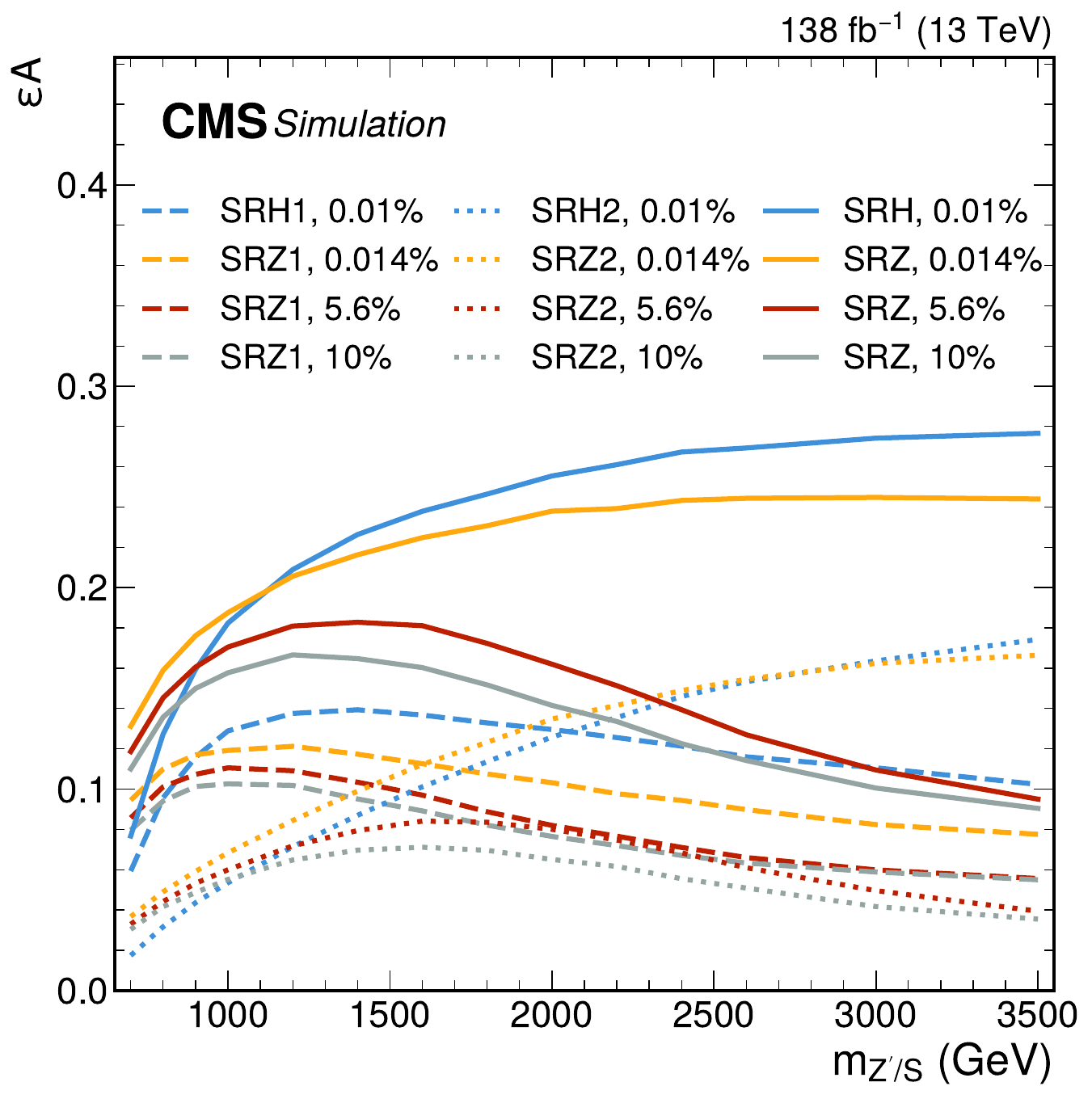}
\caption{A schematic illustration of the SR and CR definitions in the \mj-Xbb plane is shown on the left.
On the right, the product of signal efficiency and acceptance is plotted as a function of the resonance mass for the simulated signal samples in the relevant SRs.
The percentages in the legend refer to the corresponding $\Gamma/m$ values.
Here, SRZ(H) denotes SRZ(H)1+SRZ(H)2. }
\label{fig:Regions}
\end{figure}

The \bbbar jet candidates from \PH and \PZ boson decays exhibit distinctive substructures compared to background jets, which are predominantly produced in QCD processes.
This difference arises primarily from their two-pronged structure and the presence of secondary vertices from \PQb hadron decays.
These features are effectively leveraged by the \textsc{GloParT} algorithm through its dedicated Xbb tagger, which reliably assigns scores close to 1 for signal jets and close to 0 for background jets, as shown in the left plot of Fig.~\ref{fig:TaggerPlots}.
The SR selection criteria include the requirement that $\text{Xbb}>0.8$, which corresponds to a selection efficiency of approximately 68\% (0.6)\% for signal (background) jets, evaluated using the 1\TeV mass \PZpr resonance sample and the simulated background samples.
Events meeting this criterion are further split into two categories based on the Xbb score---the first SR (SR1) requires $0.8<\text{Xbb}<0.98$, and the second SR (SR2) requires $0.98<\text{Xbb}<1$. 
The SR2 has high signal purity and provides most of the sensitivity, while SR1, with moderate purity, offers a complementary gain, especially at lower resonance masses where SR2 alone is less efficient.
The Xbb boundary values of 0.8 and 0.98 were found to maximize the $S/\sqrt{D}$ across nearly all considered signal scenarios.
The definitions of the SRs and the control regions (CRs) in the \mj-Xbb phase space are illustrated on the left of Fig.~\ref{fig:Regions}.

The SR requires \bbbar candidate jets to have a mass in the range $80<\mj<150\GeV$ to ensure compatibility with the nominal \PZ and \PH boson masses.
Events are further categorized according to the value of \mj: those with \mj in the range 80--110\GeV are classified into the \PZ boson signal region (SRZ), while those in the 110--150\GeV range are classified into the Higgs boson signal region (SRH).
This \mj-based categorization is applied in conjunction with the Xbb-based classification, resulting in four SRs, referred to as SRZ1, SRZ2, SRH1, and SRH2. 
The two regions labeled SRZ1 and SRZ2 target the $\PS\to\PZ\PGg$ signal, while SRH1 and SRH2 target the $\PZpr\to\PH\PGg$ signal.
The product of signal selection efficiency and acceptance for each SR and signal scenario is shown on the right of Fig.~\ref{fig:Regions}.

The values that define the boundaries of SRZ and SRH are chosen such that the nominal masses of the \PH and \PZ bosons are not perfectly centered within the range. This decision is based on two factors. 
First, studies of the simulated signal samples indicate that their \mj distribution peaks approximately 4\% higher than the nominal mass, primarily as a result of additional radiation captured within the jet cone. Second, the \mj distributions of the background MC samples are falling over the relevant \mj range, resulting in a larger background contribution to the SR when the lower boundary is chosen to be symmetric with the upper boundary, relative to the nominal \PH or \PZ masses. 
These effects are illustrated in the left plot of Fig.~\ref{fig:PreselectionPlots}.
Optimization studies demonstrate that the selected SRZ and SRH jet mass boundaries offer optimal sensitivity across all signal masses considered. 
All kinematic boundaries defining the SRs are chosen to maximize the $S/\sqrt{D}$ metric for both SRs and preselection event selections.

To define a CR with kinematic features similar to the SR, but expected to receive negligible contributions from possible signal processes, the CR selection criteria apply the same requirements as the SR, with the exception that the \mj requirement is inverted. Specifically, \mj is required to lie outside the SRH or SHZ ranges, $50<\mj<80\GeV$ or $\mj>150\GeV$. 
This CR is then divided into two subregions, CR1 and CR2, based on the Xbb score, following the same categorization scheme as the SR.
The CRs are not directly used in the statistical analysis, but are employed for optimization and validation studies, as discussed in Section \ref{sec:bkg_estimation}.

The statistical analysis is performed using the \mjg distributions in the four SRs (SRZ1, SRZ2, SRH1, and SRH2), as detailed in Section~\ref{sec:bkg_estimation}, while the CRs, though not used directly in the fit, are employed for optimization and validation.

\subsection{Tagger calibration} \label{sec:calibration}

In order to account for any differences in the performance of the Xbb tagger between the data and MC, a calibration procedure is carried out to derive compensating scale factors (SFs). This calibration is undertaken for each of the SR1 and SR2 regions using a dedicated data sample.
The calibration of $\PZ,\PH\to\bbbar$ signal jets is performed using a dedicated sample of SM gluon-splitting jets, $\Pg\to\bbbar$, which mimic signal jets and serve as proxies for them. 

A boosted decision tree (BDT) classifier is used to select a high-purity sample of $\Pg\to\bbbar$ jets from a QCD multijet-dominated event sample, based on substructure properties similar to those of the $\PZ,\PH\to\bbbar$ signal jets targeted for calibration.
The SFs are derived by comparing the BDT-selected events in simulation to those in the data. 
These are applied to correct the jet selection efficiency in signal events. The derived SF values are binned by data-taking year, \PH candidate jet \pt (with the boundaries at 250, 400, and 600\GeV), and Xbb score matching the SRs (with boundaries at 0.8 and 0.98).
The SF values range from 0.94 to 1.05 for SR1, and from 1.07 to 1.22 for SR2, with associated uncertainties of 2--12\%. 
The same calibration method has been applied in several analyses~\cite{CMS:2019hve,CMS:2022gjd,CMS:2024fkb}. 
Details for the calibration of \ccbar and \bbbar jet tagging, and the determination of their corresponding uncertainties, can be found in Refs.~\cite{CMS-DP-2022-005,CMS:2025kje}.

\section{Background modeling} \label{sec:bkg_estimation}

Based on MC studies, it is observed that the background in SRZ1 and SRZ2 primarily arises from nonresonant SM QCD+$\PGg$ and Z+$\PGg$ processes (collectively about 96\%), which dominate the low and high regions of the \mjg spectra, respectively.
Processes such as $\ttbar$+\PGg, single $\PQt$+$\PGg$, $\PW$+$\PGg$, as well as those involving misidentified photons, are highly suppressed, contributing to only about 4\% of the selected events.
Similarly, the background in the SRH1 and SRH2 regions consists predominantly of events from QCD+$\PGg$ (about 84\%), followed by $\PZ$+$\PGg$ processes (10\%), with the remaining 6\% attributed to various other SM processes.

To avoid MC limitations and potential residual mismodeling, an approach based on data is used to model the background. 
This technique relies on creating the SM background prediction by directly fitting a set of smooth, monotonically falling functions to the \mjg spectra in the data, independently in each of the four SRs. 

A set of six functional forms of the \mjg variable is considered for performing this background fit and is employed in the statistical analysis:
\begin{equation} \begin{aligned} \label{eq:BKG}
f_1(m) & = p_0 m^{p_1+p_2\log(m)}&(\text{Dijet2}),\\
f_2(m) & = p_0 m^{p_1+p_2\log(m)+p_3\log^2(m)}&(\text{Dijet3}),\\
f_3(m) & = p_0 m^{p_1}&(\text{ExPow1}),\\
f_4(m) & = p_0 m^{p_1} e^{p_2m}&(\text{ExPow2}),\\
f_5(m) & = p_0 (1+p_1m)^{p_2}&(\text{InvPow2}),\\
f_6(m) & = p_0 (1+p_1m)^{p_2+p_3m}&(\text{InvPow3}),
\end{aligned} \end{equation}
where $m$ is defined as $m=\mjg/\sqrt{s}$ and $p_i$ are parameters constrained in the fit.
The signal strength is extracted through an unbinned maximum likelihood fit using a profile likelihood test statistic~\cite{CMS:2024onh}. 
To metigate uncertainty and bias arising from the choice of functional form, a discrete profiling technique~\cite{Dauncey:2014xga,CMS:2014afl,ATLAS:2015yey} is employed. 
In this approach, the fit function is treated as a discrete nuisance parameter and profiled simultaneously with the continuous ones, to determine the optimal combination of background models for each signal mass hypothesis. This enables different functions, or combinations of those provided, to be selected in different regions of the likelihood parameter space, mitigating the bias from using a single functional form across the entire fit range.
 
\begin{figure}[!htbp] \centering
\includegraphics[width=0.495\linewidth]{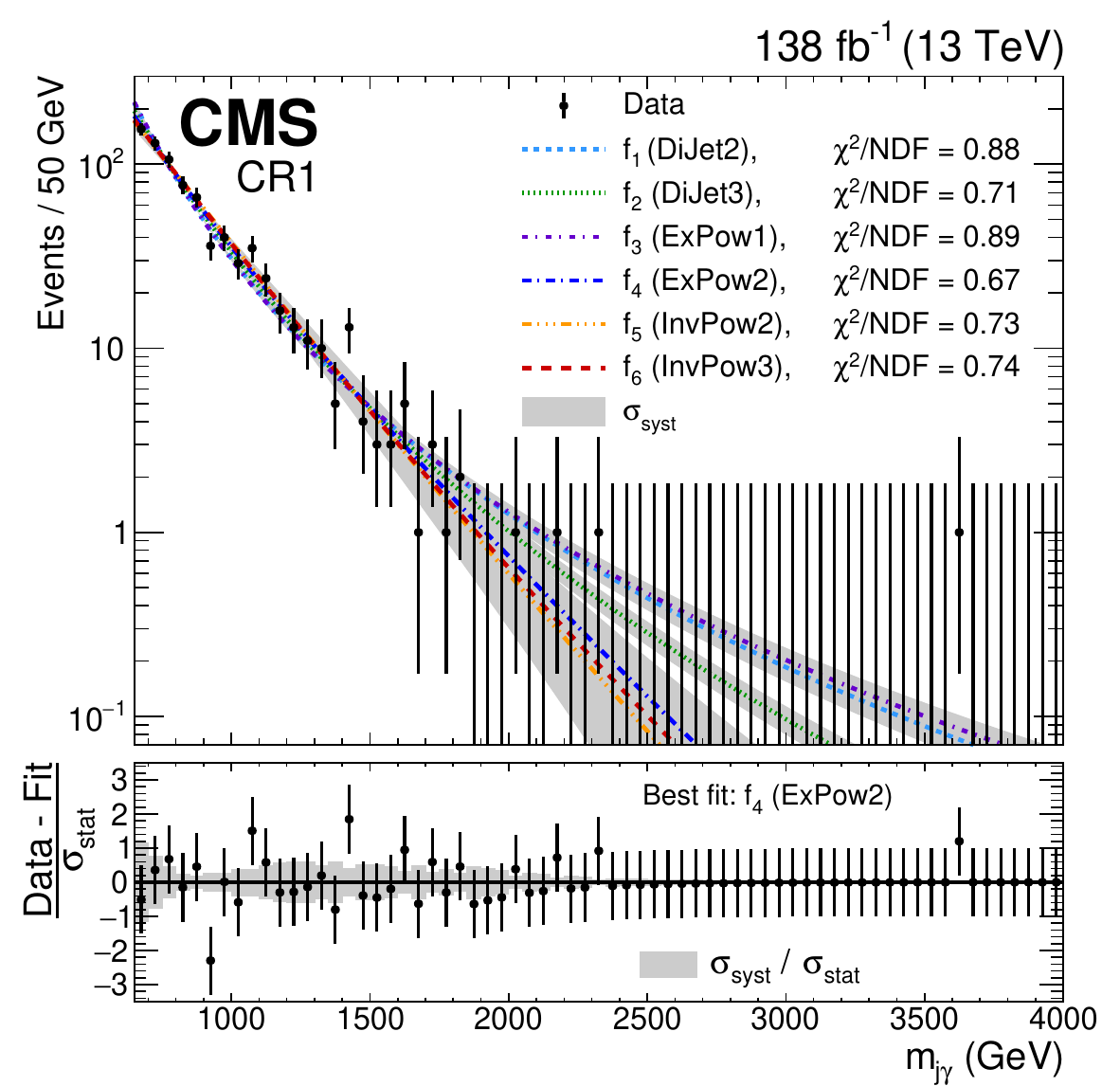}
\includegraphics[width=0.495\linewidth]{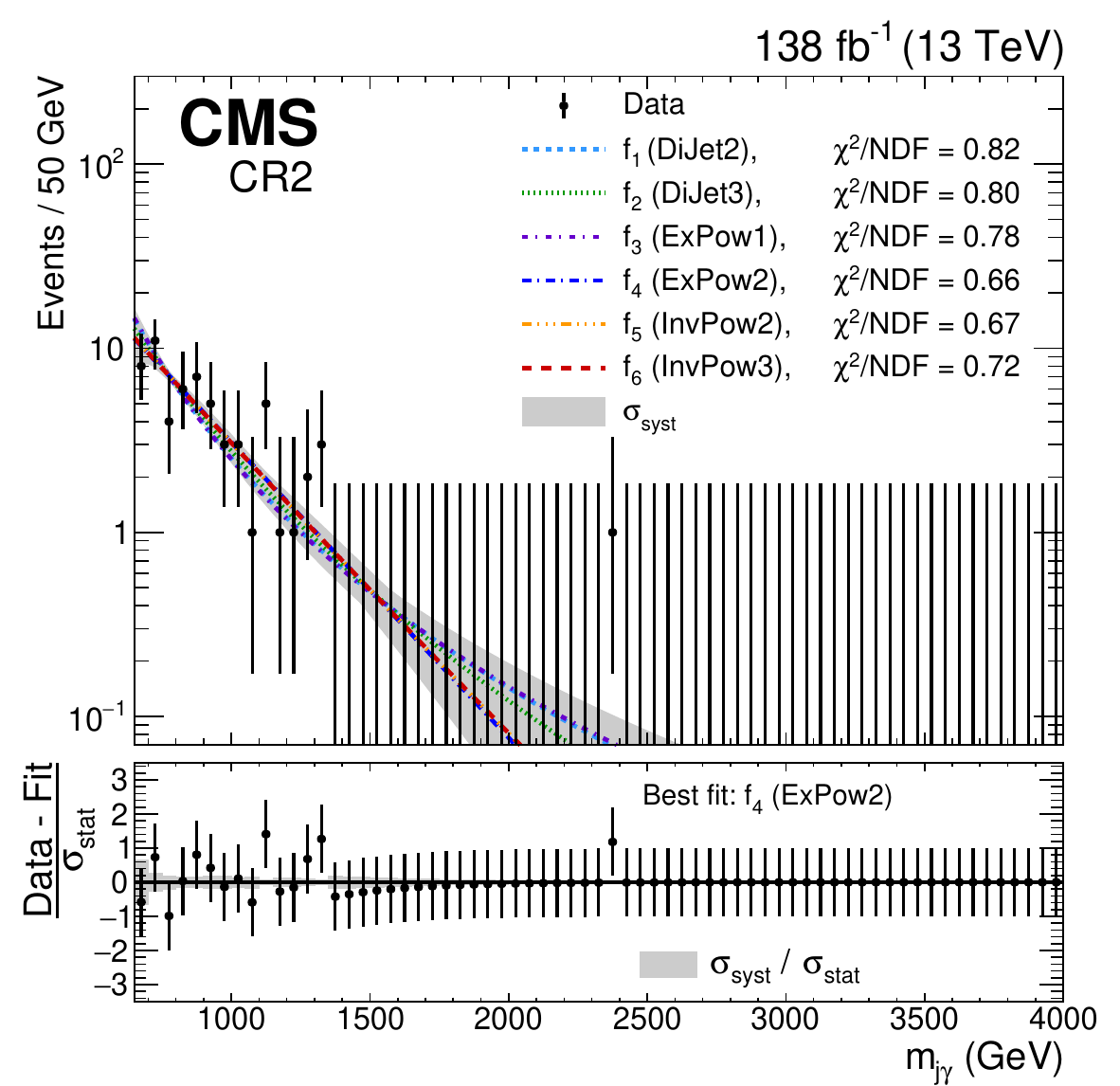}
\caption{The \mjg distributions of the data in CR1 (left) and CR2 (right) fitted with the six parametric functions. 
The legends specify the $\chi^2/\text{ndf}$ for each fit.
The lower panels show the pull distributions as defined in the text, with respect to the best fit function specified in the legend.}
\label{fig:Results}
\end{figure}

The functional forms chosen for the background fit have been employed in numerous analyses involving two final-state physics objects, including dijet, diphoton, and photon+jet systems (\eg,~\cite{CMS:2017dyb,CMS:2018cno,CMS:2021zxu,CMS:2018dqv,CMS:2024nht,CMS:2022usq}).
These families of functions are validated using data in the CRs, as well as the MC background samples in both the CRs and SRs. These validation studies consistently demonstrate a good description of the observed \mjg spectra in all cases.
The plots on the left and right of Fig.~\ref{fig:Results} show the \mjg spectra of the data in CR1 and CR2, respectively, together with the fit results. Each plot also shows the corresponding $\chi^2/\mathrm{ndf}$ value (where the ndf is the number of degrees of freedom, ranging 63--65) used to quantify the fit quality.
The indicated uncertainties on the data, $\sigma_{\text{stat}}$, represent Poisson statistical uncertainties.
The systematic uncertainty shown, $\sigma_{\text{syst}}$, corresponds to an overlay of all six fit functions' statistical uncertainties, while the  $\sigma_{\text{stat}}$ is obtained directly from the data. 
The lower panels show the pull distributions, defined as the difference between the observed and predicted event yields from the best-fit function, divided by $\sigma_{\text{stat}}$. They also display the ratios of $\sigma_{\text{syst}}$ to $\sigma_{\text{stat}}$.

The potential bias on the estimated signal significance arising from the chosen background parameterization is tested following the approach of Refs.~\cite{CMS:2018cno,CMS:2018dqv,Tsoi:2024pbn}, by generating pseudo-datasets and computing significances with the background function either fixed to that used in the generation or allowed to float in the fit. In all cases, no significant bias is induced.

\section{Systematic uncertainties} \label{sec:uncertainties}

Systematic uncertainties are considered for both background estimation and signal extraction.
Each source of uncertainty is modeled with a nuisance parameter.
These are listed in Table~\ref{tab:SysUnc}.
The table also specifies whether an uncertainty affects the shape of the \mjg spectrum or the rate of a given process.
The potential bias from using only six functional forms in the fit is negligible, based on MC studies and results from similar analyses~\cite{CMS:2017dyb,CMS:2018cno,CMS:2021zxu,CMS:2024nht}.

\begin{table}[!htbp]\centering
\topcaption{The sources of systematic uncertainties included in the analysis.
The second column indicates whether the uncertainty affects the background or signal shape or its rate. The third column from the left lists the magnitude of the corresponding pre-fit systematic uncertainty. The last column indicates the total number of nuisance parameters (NPs) and whether or not they are treated as correlated across all SRs.
An asterisk (*) denotes a value or shape template unique to each signal scenario.}
\cmsTable{
\begin{tabular}{l c c c}
Uncertainty source & Effect on & Magnitude & Number of NPs \& correlations\\ 
[\cmsTabSkip]
\hline \\[-10pt]
Background function choice & Shape & \NA & 1 per SR, uncorr. across SRs\\
[\cmsTabSkip]
Photon identification SFs & Rate & 4.4\% & 1, correlated across all SRs\\
Trigger selection SFs & Rate & 1\% & 1, correlated across all SRs\\
Integrated luminosity & Rate & 1.6\% & 1, correlated across all SRs\\
PU reweighting & Rate & 1\% & 1, correlated across all SRs\\
PDFs & Rate & 0.3--5.0\% & 1, correlated across all SRs *\\
$\mu_\text{R}$ and $\mu_\text{F}$ scales & Rate & 0.2--1.2\% & 1, correlated across all SRs *\\
Xbb tagger SF systematic & Rate & 3--7\%  & 2, uncorr. across SRs *\\ 
Xbb tagger SF statistical & Rate & \textless 4\% & 6, uncorr. across years and SRs *\\ [\cmsTabSkip]
JES & Shape &  0.7--4.0\% on the peak position & 1, per data-taking year *\\
JER & Shape & 0.5--4.0\% on the peak width & 1, correlated across SRs *\\
PES & Shape & 0.3--0.7\% on the peak position & 1, correlated across SRs \\
PER & Shape & 0.3--0.8\% on the peak width & 1, correlated across SRs
\end{tabular} }
\label{tab:SysUnc}
\end{table}

\subsection{Systematic uncertainties in the background estimation} 

As described in Section~\ref{sec:bkg_estimation}, the uncertainty associated with the choice of the background fit function is accounted for in the discrete profiling method by treating the choice of the background fit function as a discrete nuisance parameter, which is profiled in a manner analogous to the continuous nuisance parameters in the statistical analysis.
This approach minimizes potential bias in the fit function by relying on data to evaluate both the rate and shape of the predicted spectra. 

\subsection{Systematic uncertainties in the signal rate}

The uncertainty in the integrated luminosity for the 2016, 2017, and 2018 data-taking periods is 1.2--2.5\%~\cite{CMS-LUM-17-003,CMS-PAS-LUM-17-004,CMS-PAS-LUM-18-002}, with an overall uncertainty of 1.6\% for the entire 2016--2018 period.
Uncertainty introduced by the PU reweighting procedure is assessed by varying the total inelastic cross section (assumed to be 69.2 mb) by $\pm4.6$\%~\cite{Sirunyan:2018nqx}, leading to a 1\% uncertainty in signal normalization in the SRs. It is observed that this variation has a negligible effect on the shape of the signals' \mjg distributions.
Both the integrated luminosity uncertainty and the PU reweighting uncertainty are incorporated in the statistical analysis as nuisance parameters of log-normal type~\cite{CMS:2024onh}, hereafter referred to as log-normal priors.

Uncertainties of 4.4 and 1.0\% from photon identification and trigger efficiency, respectively, are included based on data-to-simulation comparisons. They are treated as correlated across all signal regions and data-taking years, and are modeled with log-normal prior nuisance parameters.

The renormalization ($\mu_\text{R}$) and factorization ($\mu_\text{F}$) scale uncertainties, along with PDF uncertainties affecting signal acceptance, are evaluated independently for each SR and signal scenario.
Scale uncertainties are derived by varying $\mu_\text{R}$ and $\mu_\text{F}$ independently by factors of 1/2 and 2, excluding extreme opposite-direction variations. 
Half of the maximum difference from these variations is assigned as the uncertainty, modeled with a log-normal prior, resulting in a $<$1.2\% effect on the signal rate.
The PDF uncertainties are evaluated using replicas of the NNPDF set and are applied to the selection acceptance for each signal scenario.
Elements of the Hessian matrix are varied, and the quadratic sum of deviations from the central value is computed for each SR. 
The resulting variation in signal yields ranges from 0.3--5.1\% across all signals considered.
One nuisance parameter accounts for combined $\mu_\text{R}$ and $\mu_\text{F}$ uncertainties, and another for PDF variations. 
Both are treated with log-normal priors.

\subsection{Systematic uncertainties in the signal shape}

The JES uncertainty is assessed by varying the reconstructed four-momenta of all jets in an event based on their $\eta$- and \pt-dependent uncertainties. 
The resulting effect on the \mjg distributions is taken as the uncertainty.
To account for differences in JER between data and simulation, smearing is applied to the latter, with jet resolutions varied according to their uncertainties.
Both sources of uncertainty are modeled using nuisance parameters affecting the signals' \mjg spectra, with Gaussian priors.
The effects of the photon energy scale (PES) and resolution (PER) are treated as uncertainties affecting only the signal shape, as their effects on signal rates were found to be negligible. These effects are considerably less significant than those arising from JES and JER due to the higher precision of photon energy reconstruction~\cite{CMS:2020uim}.

\subsection{Systematic uncertainties in the Xbb tagger selection}

Uncertainties in the Xbb tagger calibration, \ie, the derived SFs, are evaluated and summarized in Table~\ref{tab:SysUnc}.
The dominant one arises from comparing SFs derived using two reweighting variations addressing mismodelling in the BDT score and the fit variable with the nominal result~\cite{CMS-DP-2022-005,CMS:2025kje}.
The associated SF systematic uncertainty ranges from 3--7\% for \PZ and \PH boson candidate jets, accounting for residual jet shower mismodeling.
A different nuisance parameter is assigned for SR1 and SR2, each correlated across all data-taking years.
Statistical uncertainties in the SFs, arising from limited sample sizes, 1--4\%, are treated with separate nuisances per SR and data-taking year.
A log-normal prior is used for all nuisance parameters.
The impact on the \mjg shape is found to be negligible.

\section{Statistical analysis and results} \label{sec:Results}

\begin{figure}[!htbp]\centering
\includegraphics[width=0.495\linewidth]{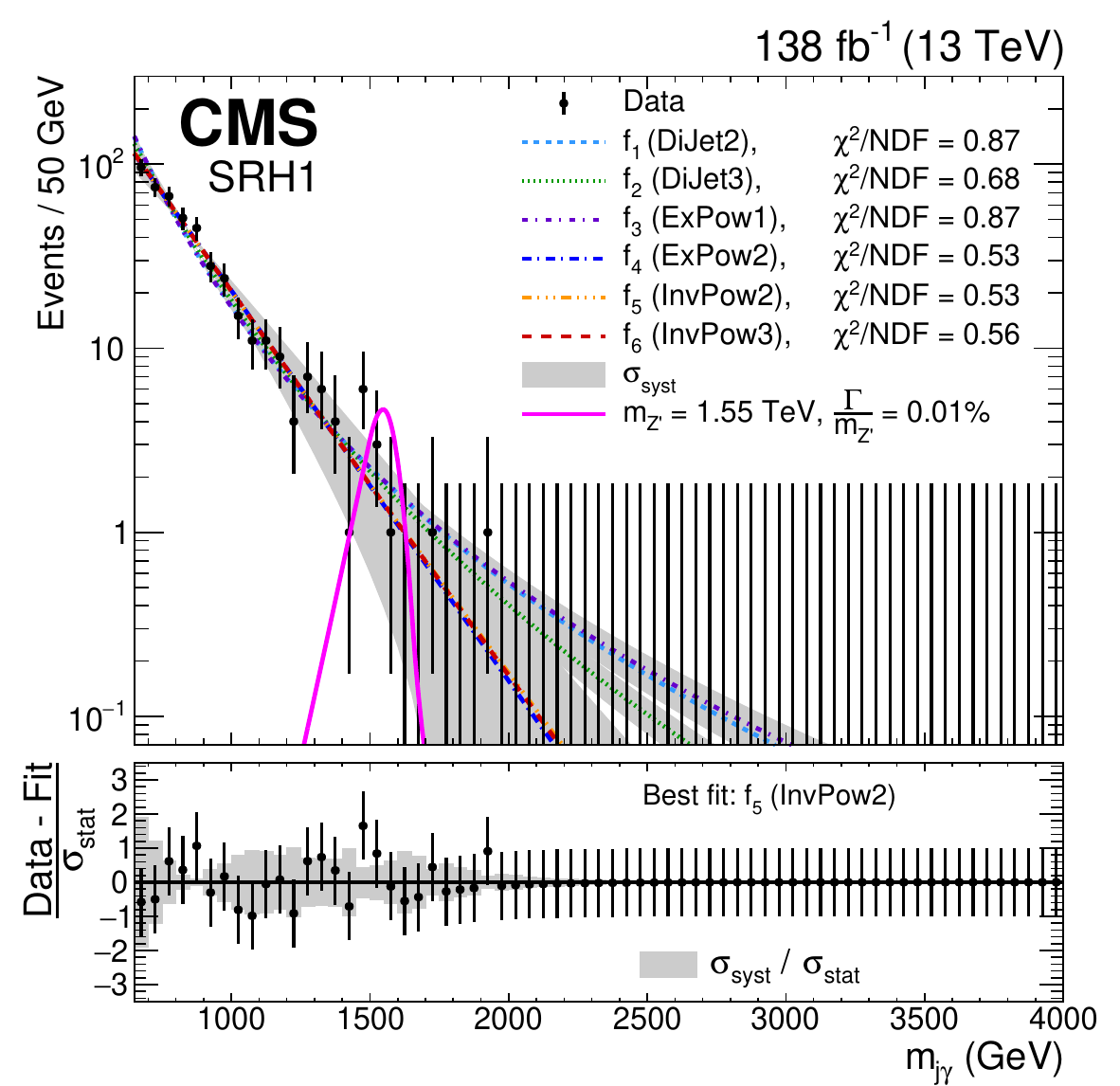}
\includegraphics[width=0.495\linewidth]{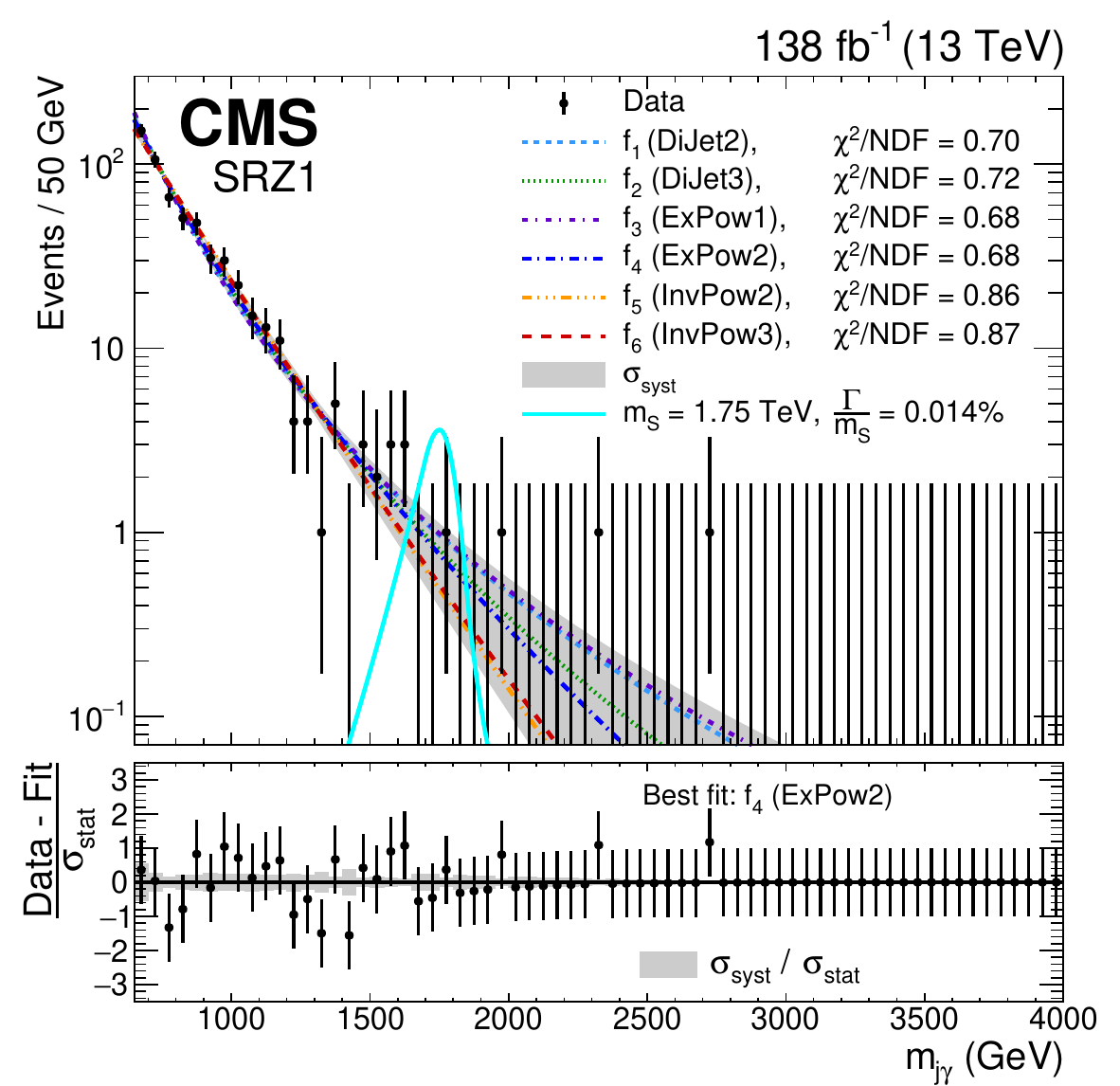}
\includegraphics[width=0.495\linewidth]{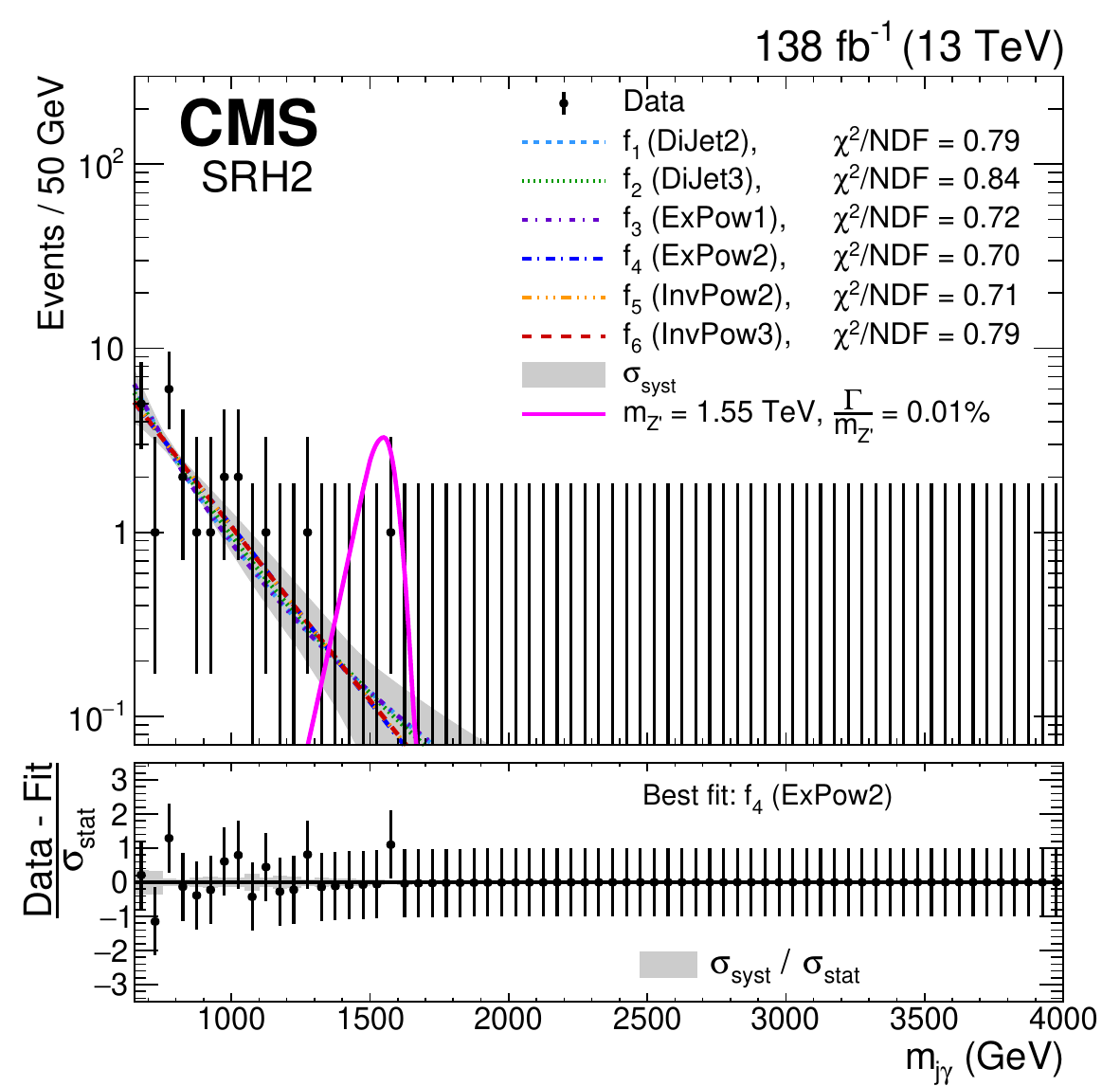}
\includegraphics[width=0.495\linewidth]{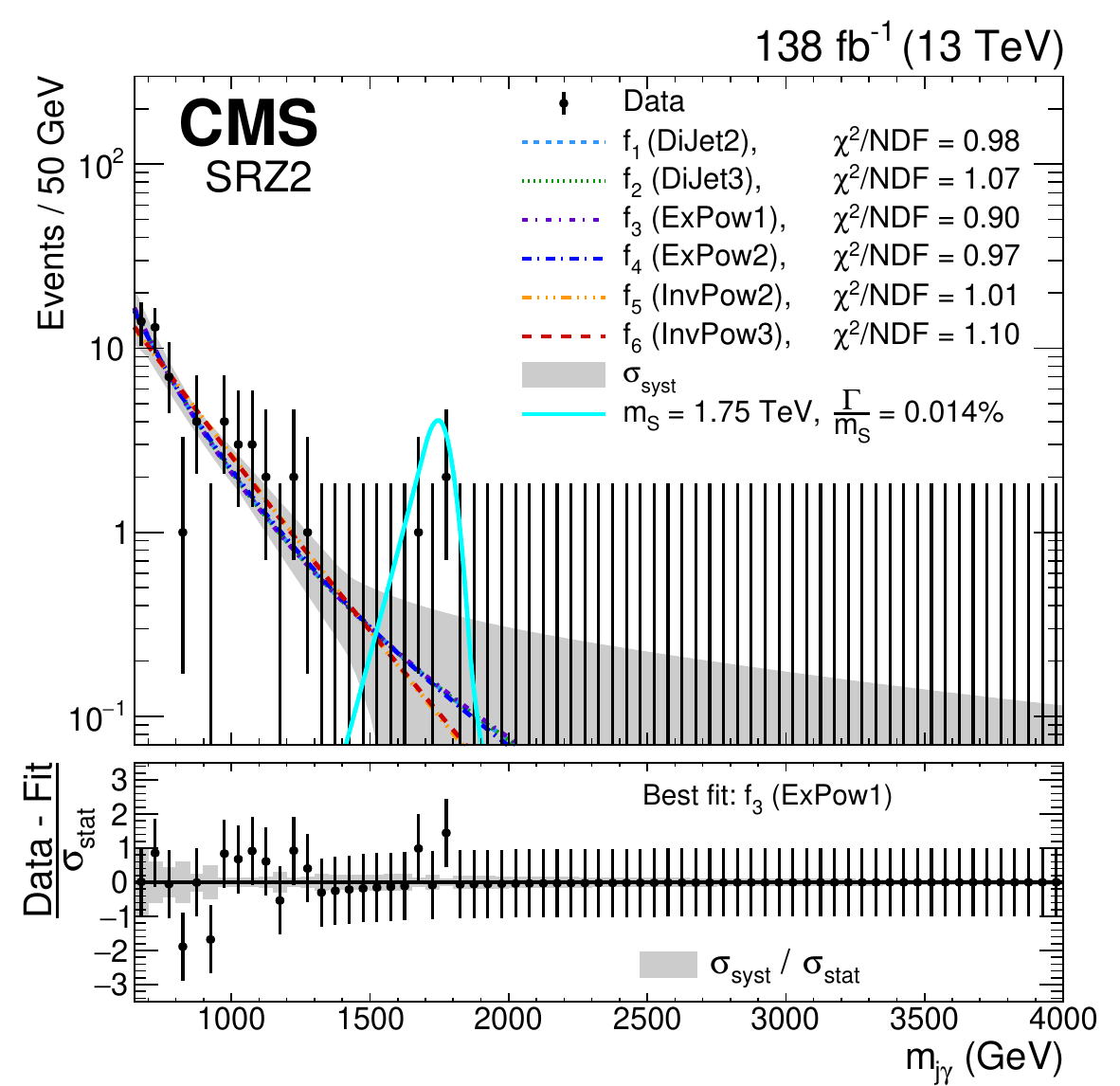}
\caption{Post-fit \mjg spectra in the four SRs:
 SRH1 (upper left), SRZ1 (upper right), SRH2 (lower left), and SRZ2 (lower right).
The legends specify the $\chi^2/\text{ndf}$ for each fit.
The lower panels show the pull distributions with respect to the best fit function specified in the legend.
The signals with the largest local significances are shown with continuous lines in each SR and are normalized to the observed $\sigma\mathcal{B}$ upper limits.}
\label{fig:Postfit}
\end{figure}

As described in Section~\ref{sec:bkg_estimation}, the search for \PZpr and \PS resonances is conducted using an unbinned maximum likelihood fit to the \mjg distribution across all four SRs.
The results, determined using the CMS statistical analysis tool \textsc{Combine}~\cite{CMS:2024onh}, are shown in Fig.~\ref{fig:Postfit}, where for illustration purposes the data are shown with 50\GeV binning.
No data events are observed beyond the displayed \mjg range.
The data agree with the SM background prediction fit functions within the uncertainties.

The uncertainty in the background model arises from the limited number of data events at high \mjg, which affects both the precision of the fit parameters and the choice of functional form.
For the signal model, the dominant sources of systematic uncertainty are the Xbb tagger SF calibration and the JES.
In the lower panels of Fig.~\ref{fig:Postfit}, the ratio of systematic to statistical uncertainties, $\sigma_{\text{syst}}/\sigma_{\text{stat}}$, indicates the ranges where the uncertainty is dominated by the functional forms used in the fit or by the observed number of data events.

\begin{figure}[!htbp]\centering
\includegraphics[width=0.495\linewidth]{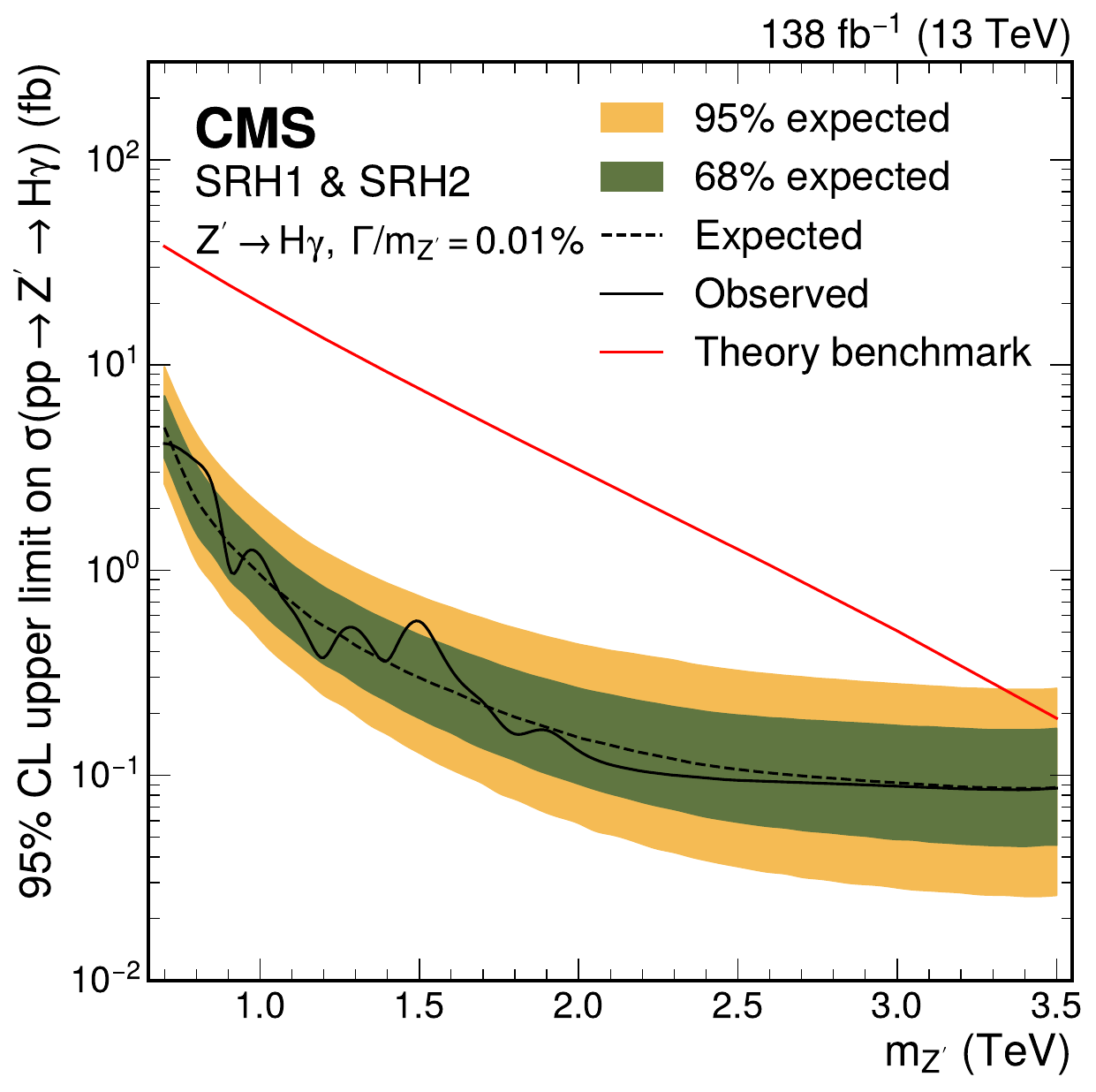}
\includegraphics[width=0.495\linewidth]{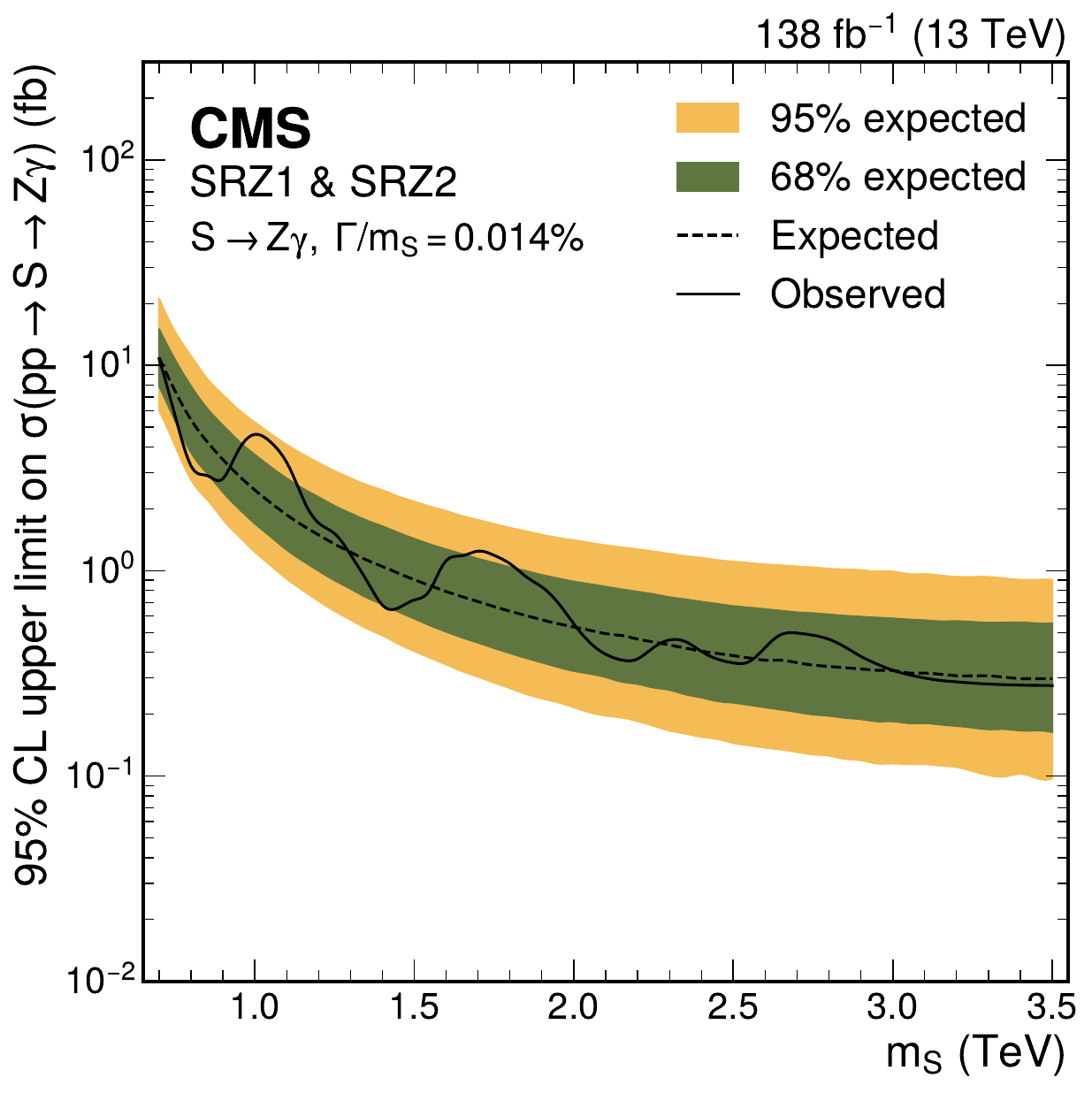}
\includegraphics[width=0.495\linewidth]{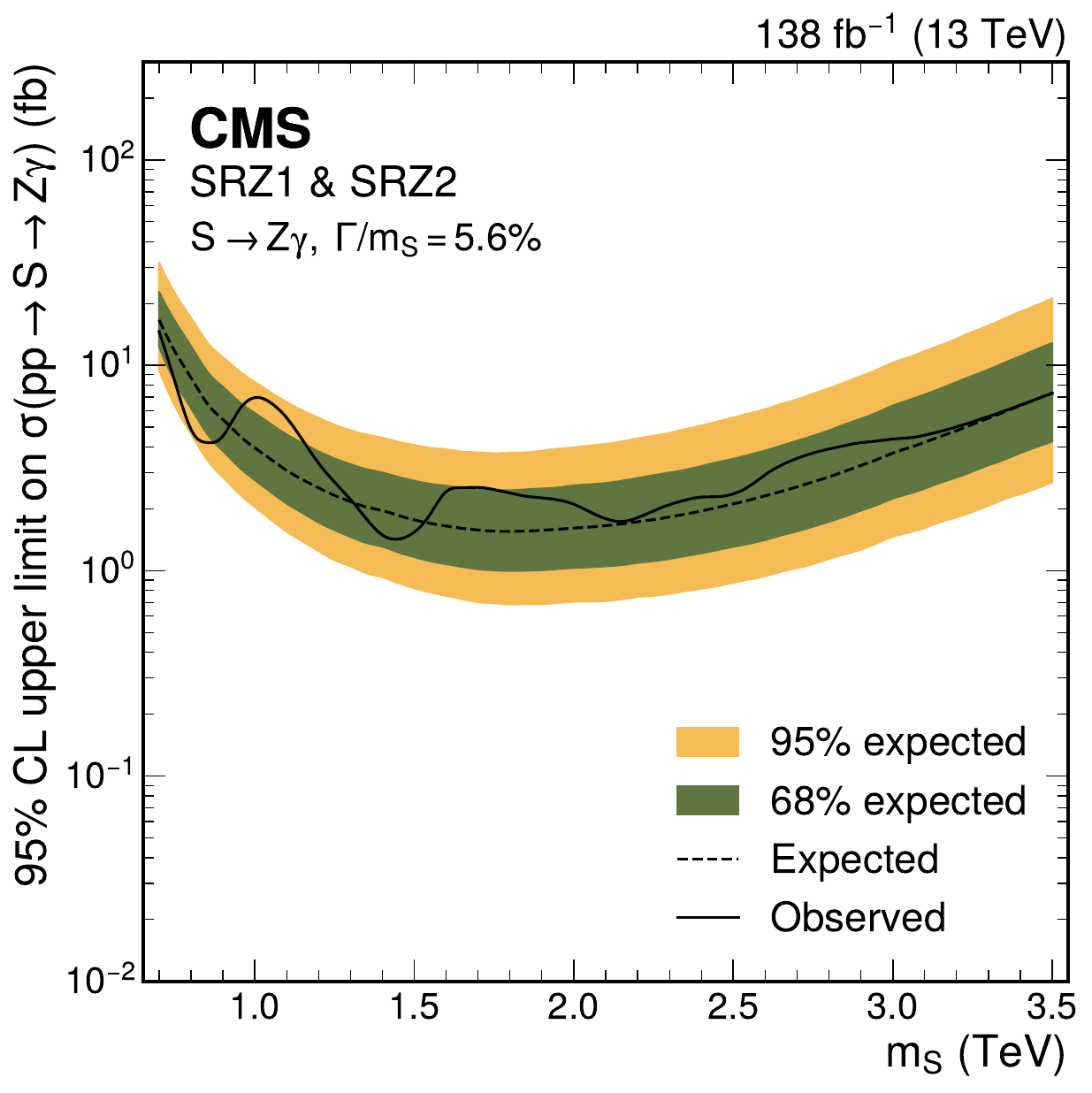}
\includegraphics[width=0.495\linewidth]{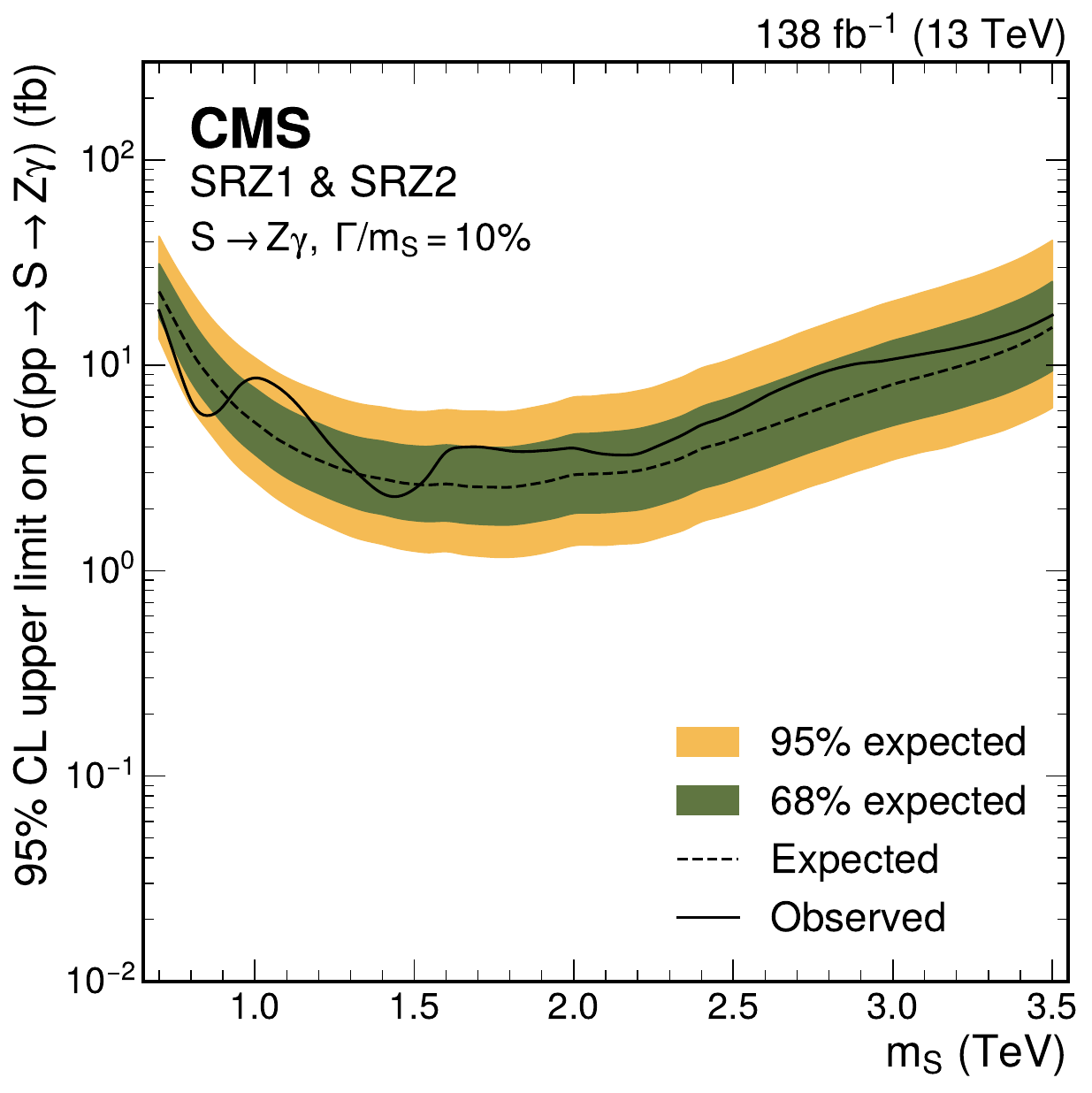}
\caption{The 95\% \CL upper limits on the product of production cross section and branching fraction $\sigma\mathcal{B}$ for $\PZpr\to\PH\PGg$ (upper left) and $\PS\to\PZ\PGg$ with a narrow width (upper right), 5.6\% width (lower left), and 10\% width (lower right).
Observed (expected) limits are shown with solid (dashed) lines.
The colored bands represent the 68 and 95\% \CL intervals for the expected limits.
The red line represents the theory benchmark model used for \PZpr signal simulation.
}
\label{fig:Limits}
\end{figure}

Upper limits at 95\% confidence level (\CL) are set on the product of the cross section and branching fraction ($\sigma\mathcal{B}$) for potential resonance signals as functions of \mZp~and \mS, shown in Fig.~\ref{fig:Limits}.
The limits are derived using the \CLs criterion~\cite{CLS2,CLS1}, based on a profile likelihood ratio test statistic in the LHC-style convention~\cite{CMS:2024onh}, modified for upper limit setting and evaluated using the asymptotic approximation~\cite{Cowan:2010js}.

For large widths, the limits weaken at high masses because a significant fraction of signal events are produced off-shell, as illustrated in the right panel of Fig.~\ref{fig:PreselectionPlots}, leading to softer photons or jets than those selected. 
Furthermore, in the broad-width scenarios, only the on-shell component of the \mjg distribution within a window of $\pm0.33 \mS$ around the nominal resonance mass is considered, where the DCB model remains valid.

This search excludes, at 95\% \CL, \PZpr resonances decaying to \PH and a photon, with \mZp in the range of 0.7--3.5\TeV, based on the benchmark model proposed in Ref.~\cite{Dobrescu:2017sue}. 
The observed limits on $\sigma\mathcal{B}$ set by this analysis are the most stringent to date for the benchmark \PZpr model across the full mass range probed, surpassing previous best results from ATLAS~\cite{ATLAS:2020jeb} by roughly a factor of 3 and improving upon earlier CMS results~\cite{CMS:2018cno} based on 2016 data by a factor of 6--10. 
For the \PH-like scalar \PS, the observed limits provide the most stringent constraints to date on $\sigma\mathcal{B}$ for $\mS>1.2\TeV$, surpassing those reported by ATLAS in Ref.~\cite{ATLAS:2023wqy}.

\section{Summary} \label{sec:sum}

A search for heavy resonances decaying to a photon and a \PZ or a Higgs boson (\PH) in the $\PGg$+jet final state, using $\sqrt{s} = 13\TeV$ proton-proton collision data collected with the CMS detector in 2016--2018, corresponding to an integrated luminosity of 138 fb$^{-1}$, has been presented.
For the $\PH\PGg$ resonance analysis, a benchmark spin-1 resonance model with a narrow width is considered, while the $\PZ\PGg$ analysis considers a standard model Higgs-like heavy spin-0 resonance, using several different width hypotheses. 
The final states of these resonant processes feature a photon and a massive, large-radius jet, containing the decay products of $\PH,\PZ\to\bbbar$, identified using the particle transformer jet substructure algorithm \textsc{GloParT} for jet classification and the \textsc{ParticleNet} algorithm for jet mass regression. These advanced machine-learning techniques greatly improve sensitivity over previous searches.
The results are consistent with the predictions of the standard model within the measurement uncertainties.
Exclusion limits at 95\% confidence level are set on the product of the production cross section and the branching fraction for the resonance decay into $\PH\PGg$ or $\PZ\PGg$, with observed values below 0.1 and 0.3 fb for the spin-1 and spin-0 scenarios, respectively.
This result establishes the most stringent constraints to date on the production and relevant decay of such heavy resonances. 

\begin{acknowledgments}
We congratulate our colleagues in the CERN accelerator departments for the excellent performance of the LHC and thank the technical and administrative staffs at CERN and at other CMS institutes for their contributions to the success of the CMS effort. In addition, we gratefully acknowledge the computing centers and personnel of the Worldwide LHC Computing Grid and other centers for delivering so effectively the computing infrastructure essential to our analyses. Finally, we acknowledge the enduring support for the construction and operation of the LHC, the CMS detector, and the supporting computing infrastructure provided by the following funding agencies: SC (Armenia), BMBWF and FWF (Austria); FNRS and FWO (Belgium); CNPq, CAPES, FAPERJ, FAPERGS, and FAPESP (Brazil); MES and BNSF (Bulgaria); CERN; CAS, MoST, and NSFC; MINCIENCIAS (Colombia); MSES and CSF (Croatia); RIF (Cyprus); SENESCYT (Ecuador); ERC PRG, TARISTU24-TK10 and MoER TK202 (Estonia); Academy of Finland, MEC, and HIP (Finland); CEA and CNRS/IN2P3 (France); SRNSF (Georgia); BMFTR, DFG, and HGF (Germany); GSRI (Greece); NKFIH (Hungary); DAE and DST (India); IPM (Iran); SFI (Ireland); INFN (Italy); MSIT and NRF (Republic of Korea); MES (Latvia); LMTLT (Lithuania); MOE and UM (Malaysia); BUAP, CINVESTAV, CONACYT, LNS, SEP, and UASLP-FAI (Mexico); MOS (Montenegro); MBIE (New Zealand); PAEC (Pakistan); MES, NSC, and NAWA (Poland); FCT (Portugal); MESTD (Serbia); MICIU/AEI and PCTI (Spain); MOSTR (Sri Lanka); Swiss Funding Agencies (Switzerland); MST (Taipei); MHESI (Thailand); TUBITAK and TENMAK (T\"{u}rkiye); NASU (Ukraine); STFC (United Kingdom); DOE and NSF (USA).

\hyphenation{Rachada-pisek} Individuals have received support from the Marie-Curie program and the European Research Council and Horizon 2020 Grant, contract Nos.\ 675440, 724704, 752730, 758316, 765710, 824093, 101115353, 101002207, 101001205, and COST Action CA16108 (European Union); the Leventis Foundation; the Alfred P.\ Sloan Foundation; the Alexander von Humboldt Foundation; the Science Committee, project no. 22rl-037 (Armenia); the Fonds pour la Formation \`a la Recherche dans l'Industrie et dans l'Agriculture (FRIA) and Fonds voor Wetenschappelijk Onderzoek contract No. 1228724N (Belgium); the Beijing Municipal Science \& Technology Commission, No. Z191100007219010, the Fundamental Research Funds for the Central Universities, the Ministry of Science and Technology of China under Grant No. 2023YFA1605804, the Natural Science Foundation of China under Grant No. 12061141002, 12535004, and USTC Research Funds of the Double First-Class Initiative No.\ YD2030002017; the Ministry of Education, Youth and Sports (MEYS) of the Czech Republic; the Shota Rustaveli National Science Foundation, grant FR-22-985 (Georgia); the Deutsche Forschungsgemeinschaft (DFG), among others, under Germany's Excellence Strategy -- EXC 2121 ``Quantum Universe" -- 390833306, and under project number 400140256 - GRK2497; the Hellenic Foundation for Research and Innovation (HFRI), Project Number 2288 (Greece); the Hungarian Academy of Sciences, the New National Excellence Program - \'UNKP, the NKFIH research grants K 131991, K 133046, K 138136, K 143460, K 143477, K 146913, K 146914, K 147048, 2020-2.2.1-ED-2021-00181, TKP2021-NKTA-64, and 2021-4.1.2-NEMZ\_KI-2024-00036 (Hungary); the Council of Science and Industrial Research, India; ICSC -- National Research Center for High Performance Computing, Big Data and Quantum Computing, FAIR -- Future Artificial Intelligence Research, and CUP I53D23001070006 (Mission 4 Component 1), funded by the NextGenerationEU program (Italy); the Latvian Council of Science; the Ministry of Education and Science, project no. 2022/WK/14, and the National Science Center, contracts Opus 2021/41/B/ST2/01369, 2021/43/B/ST2/01552, 2023/49/B/ST2/03273, and the NAWA contract BPN/PPO/2021/1/00011 (Poland); the Funda\c{c}\~ao para a Ci\^encia e a Tecnologia, grant CEECIND/01334/2018 (Portugal); the National Priorities Research Program by Qatar National Research Fund; MICIU/AEI/10.13039/501100011033, ERDF/EU, "European Union NextGenerationEU/PRTR", and Programa Severo Ochoa del Principado de Asturias (Spain); the Chulalongkorn Academic into Its 2nd Century Project Advancement Project, the National Science, Research and Innovation Fund program IND\_FF\_68\_369\_2300\_097, and the Program Management Unit for Human Resources \& Institutional Development, Research and Innovation, grant B39G680009 (Thailand); the Kavli Foundation; the Nvidia Corporation; the SuperMicro Corporation; the Welch Foundation, contract C-1845; and the Weston Havens Foundation (USA).
\end{acknowledgments}

\bibliography{auto_generated} 

\cleardoublepage \appendix\section{The CMS Collaboration \label{app:collab}}\begin{sloppypar}\hyphenpenalty=5000\widowpenalty=500\clubpenalty=5000\cmsinstitute{Yerevan Physics Institute, Yerevan, Armenia}
{\tolerance=6000
A.~Hayrapetyan, V.~Makarenko\cmsorcid{0000-0002-8406-8605}, A.~Tumasyan\cmsAuthorMark{1}\cmsorcid{0009-0000-0684-6742}
\par}
\cmsinstitute{Institut f\"{u}r Hochenergiephysik, Vienna, Austria}
{\tolerance=6000
W.~Adam\cmsorcid{0000-0001-9099-4341}, J.W.~Andrejkovic, L.~Benato\cmsorcid{0000-0001-5135-7489}, T.~Bergauer\cmsorcid{0000-0002-5786-0293}, M.~Dragicevic\cmsorcid{0000-0003-1967-6783}, C.~Giordano, P.S.~Hussain\cmsorcid{0000-0002-4825-5278}, M.~Jeitler\cmsAuthorMark{2}\cmsorcid{0000-0002-5141-9560}, N.~Krammer\cmsorcid{0000-0002-0548-0985}, A.~Li\cmsorcid{0000-0002-4547-116X}, D.~Liko\cmsorcid{0000-0002-3380-473X}, M.~Matthewman, I.~Mikulec\cmsorcid{0000-0003-0385-2746}, J.~Schieck\cmsAuthorMark{2}\cmsorcid{0000-0002-1058-8093}, R.~Sch\"{o}fbeck\cmsAuthorMark{2}\cmsorcid{0000-0002-2332-8784}, D.~Schwarz\cmsorcid{0000-0002-3821-7331}, M.~Shooshtari\cmsorcid{0009-0004-8882-4887}, M.~Sonawane\cmsorcid{0000-0003-0510-7010}, W.~Waltenberger\cmsorcid{0000-0002-6215-7228}, C.-E.~Wulz\cmsAuthorMark{2}\cmsorcid{0000-0001-9226-5812}
% ...
\par}
\vskip\cmsinstskip
\dag:~Deceased\\
$^{1}$Also at Yerevan State University, Yerevan, Armenia\\
$^{2}$Also at TU Wien, Vienna, Austria\\
% ...
\end{sloppypar}
\end{document}